\documentclass[usenatbib]{mn2e}

\usepackage{savesym}
\usepackage{amsmath}
\savesymbol{iint}
\usepackage{txfonts}
\restoresymbol{TXF}{iint}

\usepackage{subfigure}
\usepackage{epsfig}
\usepackage{enumerate}
\usepackage{url}
\def\simless{\mathbin{\lower 3pt\hbox
{$\rlap{\raise 5pt\hbox{$\char'074$}}\mathchar"7218$}}}   
\def\simmore{\mathbin{\lower 3pt\hbox
{$\rlap{\raise 5pt\hbox{$\char'076$}}\mathchar"7218$}}}   
\newcommand{\be}{\begin{equation}}
\newcommand{\ee}{\end{equation}}
\newcommand       \bea          {\begin{eqnarray}}
\newcommand       \eea          {\end{eqnarray}}
\newcommand       \apj          {ApJ}
\newcommand       \apjl         {ApJL}
\newcommand       \aap          {A\&A}
\newcommand       \nat          {Nature}
\newcommand       \mnras        {MNRAS}

\newcommand       \aj      {AJ}

\newcommand       \araa      {ARA\&A}

\newcommand       \pasj   {PASJ}
\newcommand      \apjs {ApJ Supplements}

\def\apss{Astrophysics \& Space Science}

\def\simlt{\mathrel{\hbox{\rlap{\hbox{\lower4pt\hbox{$\sim$}}}\hbox{$<$}}}}
\def\simgt{\mathrel{\hbox{\rlap{\hbox{\lower4pt\hbox{$\sim$}}}\hbox{$>$}}}}
\def\lesssim{\mathrel{\hbox{\rlap{\hbox{\lower4pt\hbox{$\sim$}}}\hbox{$<$}}}}

\def\gtrsim{\mathrel{\hbox{\rlap{\hbox{\lower4pt\hbox{$\sim$}}}\hbox{$>$}}}}

\title[Emission from shocks in nova outflows]{Shocks in nova outflows. I. Thermal emission.}

\author[]{Brian~D.~Metzger$^{1}\thanks{E-mail: bmetzger@phys.columbia.edu}$, Romain~Hasco\"{e}t$^{1}$,  Indrek~Vurm$^{1,4}$, Andrei~M.~Beloborodov$^{1}$,
\vspace{0.25cm}
 \\{\LARGE\rm Laura Chomiuk$^{2}$, J.~L.~Sokoloski$^{1}$, Thomas Nelson$^{3}$}\\
$^{1}$Columbia Astrophysics Laboratory, Columbia University, New York, NY, 10027, USA\\ 
$^{2}$Department of Physics and Astronomy, Michigan State University, East Lansing, MI 48824, USA\\
$^{3}$School of Physics and Astronomy, University of Minnesota, Minneapolis, MN 55455, USA\\
$^{4}$Tartu Observatory, T$\tilde{o}$ravere, Tartumaa EE-61602, Estonia\\ }

\begin{document}
\date{Received / Accepted}
\pagerange{\pageref{firstpage}--\pageref{lastpage}} \pubyear{2013}

\maketitle

\label{firstpage}

\begin{abstract}

Growing evidence for shocks in nova outflows include (1) multiple velocity components in the optical spectra; (2) hard X-ray emission starting weeks to months after the outburst; (3) an early radio flare on timescales of months, in excess of that predicted from the freely expanding photo-ionized gas; and, perhaps most dramatically, (4) $\sim$ GeV gamma-ray emission.  We present a one dimensional model for the shock interaction between the fast nova outflow and a dense external shell (DES) and its associated thermal X-ray, optical, and radio emission.  The lower velocity DES could represent an earlier stage of mass loss from the white dwarf or ambient material not directly related to the thermonuclear runaway.  The forward shock is radiative initially when the density of shocked gas is highest, at which times radio emission originates from the dense cooling layer immediately downstream of the shock.  Our predicted radio light curve is characterized by sharper rises to maximum and later peak times at progressively lower frequencies, with a peak brightness temperature that is approximately independent of frequency.  We apply our model to the recent gamma-ray producing classical nova V1324 Sco, obtaining an adequate fit to the early radio maximum for reasonable assumptions about the fast nova outflow and assuming the DES possesses a characteristic velocity $\sim 10^{3}$ km s$^{-1}$ and mass $\sim$ few $10^{-4} M_{\odot}$; the former is consistent with the velocities of narrow line absorption systems observed previously in nova spectra, while the total ejecta mass of the DES and fast outflow is consistent with that inferred independently by modeling the late radio peak as uniformly expanding photo-ionized gas.   Rapid evolution of the early radio light curves require the DES to possess a steep outer density profile, which may indicate that the onset of mass loss from the white dwarf was rapid, providing indirect evidence that the DES was expelled as the result of the thermonuclear runaway event.  Reprocessed X-rays from the shock absorbed by the DES at early times are found to contribute significantly to the optical/UV emission, which we speculate may be responsible for the previously unexplained `plateaus' and secondary maxima in nova optical light curves.  

\end{abstract} 
  

\section{Introduction} 
\label{sec:intro}

Novae are sudden outbursts powered by runaway nuclear burning on the surface of a white dwarf accreting from a stellar binary companion (e.g.~\citealt{Gallagher&Starrfield78}; \citealt{Shore12} for a recent review).  Novae provide nearby laboratories for studying the physics of nuclear burning and accretion.  Accreting systems similar to those producing novae are also candidate progenitors of Type Ia supernovae (e.g.~\citealt{dellaValle&Livio96}; \citealt{Starrfield+04}) and other transients such a `.Ia' supernovae (e.g.~\citealt{Bildsten+07}) and accretion-induced collapse (e.g.~\citealt{Metzger+09}; \citealt{Darbha+10}).

Major open questions regarding novae include the quantity and time evolution of mass ejected by the thermonuclear outburst and its possible relationship to the immediate environment of the white dwarf or its binary companion.  Detailed hydrodynamical simulations of nova outbursts find that matter is unbound from the white dwarf in at least two distinct stages, driven by different physical processes and characterized by different mass loss rates and outflow velocities (e.g.~\citealt{Prialnik86}; \citealt{Yaron+05}; \citealt{Starrfield+09}).  The details of this evolution, however, depend sensitively on theoretical uncertainties such as the efficiency of convective mixing in the outer layers of the white dwarf following runaway nuclear burning (\citealt{Starrfield+00}; \citealt{Yaron+05}; \citealt{Casanova+11}).

Radio observations provide a useful tool for studying nova ejecta (e.g.~\citealt{Hjellming&Wade70}; \citealt{Seaquist+80}; \citealt{Hjellming87}; \citealt{Bode&Seaquist87}; \citealt{Sokoloski+08}; \citealt{Roy+12} for a recent review).  The standard scenario for nova radio emission invokes thermal radiation from freely expanding ionized ejecta of uniform temperature $\sim 10^{4}$ K (e.g.~\citealt{Seaquist&Palimaka77}; \citealt{Hjellming+79}; \citealt{Seaquist+80}; \citealt{Kwok83}; see \citealt{Seaquist&Bode08}  for a recent review).  This model predicts a radio flux $F_{\nu}$ that increases $\propto t^{2}$ at early times with an optically-thick spectrum ($\alpha = 2$, where $F_{\nu} \propto \nu^{\alpha}$), which after reaching its peak on a timescale of $\sim$ year, then decays with a flat spectrum ($\alpha = -0.1$) at late times once the ejecta have become optically thin to free-free absorption.  The simplest form of the standard model, which assumes a density profile corresponding to homologous expansion (the `Hubble flow'), provides a reasonable fit to the late radio data for most novae (\citealt{Hjellming+79}), from which total ejecta masses $\sim 10^{-4}M_{\odot}$ are typically inferred for classical novae (e.g.~\citealt{Seaquist&Bode08}). 

Although the standard model provides a relatively satisfactory picture of the late radio emission from nova, deviations from this simple picture are observed at early times.  A growing sample of nova radio light curves show a second maximum (an early `bump') on a timescale $\lesssim 100$ days after the visual peak (\citealt{Taylor+87b}; \citealt{Krauss+11}; \citealt{Chomiuk+12}; \citealt{Nelson+12}; \citealt{Weston+13}).  The high brightness temperature of this emission $\gtrsim 10^{5}$ K, and its flat spectrum relative to the standard model prediction, have supported the interpretation that it results from shock interaction between the nova ejecta and a dense external shell (DES).  The DES may represent matter ejected earlier in the nova outburst (`internal shocks'; e.g.~\citealt{Taylor+87b}; \citealt{Lloyd+96}; \citealt{Mukai&Ishida01}).  Alternatively, the DES could represent ambient material which is not directly related to the current nova eruption, such as mass loss from the binary system associated with the white dwarf accretion process (e.g.~\citealt{Williams+08}; \citealt{Williams&Mason10}).

Evidence for shocks in novae is present at other wavelengths.  In addition to broad P Cygni absorption lines originating from the fast $\sim$ few $10^{3}$ km s$^{-1}$ primary ejecta, the optical spectra of novae near maximum light also contain narrow absorption lines.  These lines originate from dense gas ahead of the primary outflow with lower velocities $\lesssim 10^{3}$ km s$^{-1}$ (e.g.~\citealt{Williams+08}; \citealt{Shore+13}).  This slow moving material, which is inferred to reside close to the white dwarf and to possess a high covering fraction, is likely to experience a subsequent collision with the fast outflow coming from behind (\citealt{Williams&Mason10}).  Other evidence for shocks in novae includes the deceleration of the ejecta inferred by comparing the high velocities of the early (pre-collision) primary nova ejecta with the lower velocities inferred from the late (post-collision) nebular emission (e.g.~\citealt{Friedjung&Duerbeck93}; \citealt{Williams&Mason10}).

Some novae produce $\gtrsim $ keV X-ray emission, with peak luminosities $L_{X} \sim 10^{34}-10^{35}$ erg s$^{-1}$ on typical timescales $\gtrsim 20-300$ days after the optical maximum  (\citealt{Lloyd+92}; \citealt{OBrien+94}; \citealt{Orio04}; \citealt{Sokoloski+06}; \citealt{Ness+07}; \citealt{Mukai+08}; \citealt{Krauss+11}).  This emission, which requires much higher temperatures than thermal emission from the white dwarf surface, has also been interpreted as being shock powered (e.g.~\citealt{Brecher+77}; \citealt{Lloyd+92}).  Given the relatively sparse X-ray observations of novae, a substantial fraction of novae may be accompanied by X-ray emission of similar luminosities to the current detections (\citealt{Mukai+08}).    

Additional dramatic evidence for shocks in novae is the recent discovery of $\sim$ GeV gamma-rays at times nearly coincident with the optical peak (\citealt{Abdo+10}).  The first gamma-ray novae occurred in the symbiotic binary V407 Cyg 2010, which appeared to favor a scenario in which the DES was the dense wind of the companion red giant (\citealt{Abdo+10}; \citealt{Vaytet+11}; \citealt{Martin&Dubus13}).  However, gamma-rays have now been detected from four ordinary classical novae (\citealt{Cheung+12}; \citealt{Cheung+13}; \citealt{Hill+13}; \citealt{Hays+13}; \citealt{Cheung&Hays13}; \citealt{Cheung&Jean13a}; \citealt{Cheung&Jean13b}), which demonstrates the presence of dense external material even in systems that are not embedded in the wind of an M giant or associated with recurrent novae.  

Observations across the electromagnetic spectrum thus indicate that shocks are common, if not ubiquitous, in the outflows of classical novae.  Theoretical models of nova shocks have been developed in previous works, but most have been applied to specific events or have been focused on the emission at specific wavelengths.  \citet{Taylor+87b}, for example, model the early radio peak of Nova Vulpeculae 1984 as being powered by the shock interaction between a high velocity outflow from the white dwarf with slower earlier ejecta.  \citet{Lloyd+92} model the X-ray emission of Nova Herculis 1991 as being shock powered, while \citet{Lloyd+96} calculate the free-free radio emission from hydrodynamical simulations of nova shocks.  \citet{Contini&Prialnik97} calculate the effects of shocks on the optical line spectra of the recurrent nova T Pyxidis.  

In this paper we present a one-dimensional model for shock interaction in novae and its resulting radiation.  Many of the above works, though ground-breaking, neglect one or more aspects of potentially important physics, such as the influence of the ionization state of the medium on the radio/X-ray opacity, or the effects of radiative shocks on the system dynamics and radio emission.  Here we attempt to include these details (if even in only a simple-minded way), in order to provide a unified picture that simultaneously connects the signatures of shocks at radio, optical and X-ray frequencies.  Our goal is to provide a flexible framework for interpreting multi-wavelength nova data in order to constrain the properties of the DES and to help elucidate its origin.  

This work (Paper I) is focused on thermal emission from the shocks, motivated by (1) its promise in explaining the qualitative features of the observed X-ray and radio emission and (2) the virtue that thermal processes can be calculated with greater confidence than non-thermal processes.  A better understanding of the mass and radial scale of the DES is also requisite to exploring non-thermal processes, such as the high energy particle acceleration necessary to produce the observed gamma-rays.  Non-thermal emission from nova shocks will be addressed in Paper II once the thermal framework is in place.  

The rest of this paper is organized as follows.  In $\S\ref{sec:overview}$ we overview the model for nova shocks and its thermal radiation.  In $\S\ref{sec:model}$ we present the details of our dynamical model of the shock-DES interaction.  In $\S\ref{sec:emission}$ we describe the resulting shock radiation at X-ray ($\S\ref{sec:Xrays}$), optical ($\S\ref{sec:optical}$), and radio ($\S\ref{sec:radio}$) frequencies.    In $\S\ref{sec:results}$ the results of our calculations are presented and compared to available data, focusing on the case of V1324 Sco.  In $\S\ref{sec:discussion}$ we discuss our results and in $\S\ref{sec:conclusions}$ we summarize our conclusions.  

\subsection{Physical Picture}
\label{sec:overview}

\begin{figure*}
\includegraphics[width=0.8\textwidth]{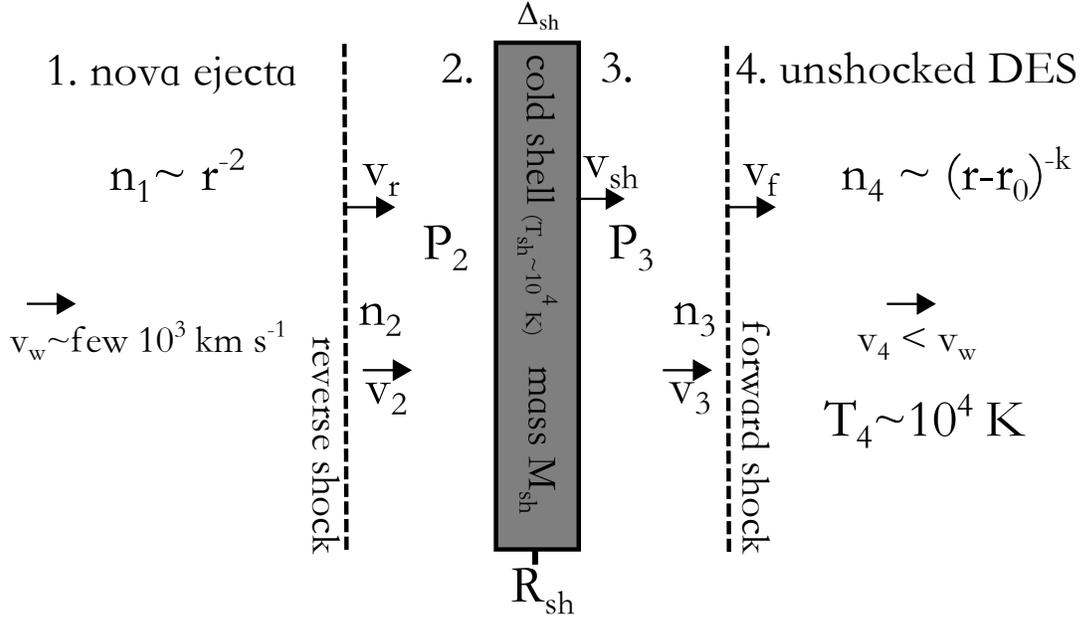}
\caption{Schematic diagram of the shock interaction between the nova outflow (Region 1) and the dense external shell (DES; Region 4).  The fast nova outflow of velocity $v_{\rm w} \sim$ few $10^{3}$ km s$^{-1}$ has a radial density profile $n_1 \propto \dot{M}_w r^{-2}$ (eq.~[\ref{eq:n1}]) characteristic of a steady wind, while the DES density scales as a power-law $n_4 \propto (r-r_0)^{-k}$ (eq.~[\ref{eq:n4}]).  A forward shock is driven into the DES, while the reverse shock is driven back into the nova ejecta.  The shocked ejecta (Region 2) and shocked DES (Region 3) are separated by a cold neutral shell containing the swept up mass of mass $M_{\rm sh}$ and thickness $\Delta_{\rm sh} \ll R_{\rm sh}$.  Emission from the reverse shock is blocked by the central shell at radio and X-ray wavelengths by free-free and bound-free absorption, respectively; optical/UV emission may also be blocked if dust forms in the shell. } 
\label{fig:schematic}
\end{figure*}

Figure \ref{fig:schematic} summarizes the physical picture.  A fast outflow from the white dwarf of velocity $v_w \sim few \times 10^{3}$ km s$^{-1}$ collides with the slower DES of velocity $v_4 < v_w$.  In this initial treatment the DES is assumed to originate from small radii (e.g. the white dwarf surface or the binary companion) starting at the time of the initial optical rise ($t = -\Delta t$, where $\Delta t$ is defined below).\footnote{An outflow starting near the optical onset is not necessarily incompatible with scenarios in which the DES represents pre-existing ambient matter unrelated to the thermonuclear runaway if the DES is accelerated by radiation pressure from the nova explosion (\citealt{Williams72}).}  The mass of the DES is concentrated about the radius $r_0 \sim v_4(t + \Delta t)$, with its density decreasing as a power-law $\propto (r-r_0)^{-k}$ at larger radii (Fig.~\ref{fig:schematic_density}).  The fast nova outflow is assumed to begin around the time of the optical maximum ($t = 0$), typically days to weeks after the optical onset.   This is justified by the fact that a substantial fraction of the optical light curve may be shock powered ($\S\ref{sec:optical}$), as evidenced in part by the coincidence between the optical peak and the peak of the gamma-ray emission in most of the {\it Fermi-}detected novae.  The fast nova outflow drives a forward shock into the DES while a reverse shock simultaneously propagates back into the nova outflow.  Both shocks are radiative at times of interest, such that the swept up gas accumulates in a cool shell between the shocks.  

Radiation observed from the shocks depends sensitively on the ionization state of the unshocked DES lying ahead.  Ionizing UV/X-ray photons produced by free-free emission at the forward shock penetrate the upstream gas to a depth that depends on the balance between photo-ionization and radiative recombination.  The structure of the resulting ionized layers controls the escape of X-rays (absorbed by neutral gas) and radio emission (absorbed by ionized gas).  At early times, when the forward shock is passing through the densest gas, X-rays are absorbed by the neutral medium ahead of the shock before reaching the observer.  A portion of the shock luminosity is re-emitted as optical/soft UV radiation, which freely escapes because of the much lower opacity at these longer wavelengths.  As we will show, this shock-heated emission may contribute appreciably to the optical/UV light curves of novae.  

At later times, as the forward shock moves to larger radii and lower densities, X-rays are able to escape, with their peak luminosity and timescale depending sensitively on the metallicity $X_Z$ of the DES.  Gas heated by the forward shock also produces radio emission originating from the dense cooling layer behind the shock.  This emission is free-free absorbed by the cooler $\sim 10^{4}$ K ionized layer just ahead of the shock (Fig.~\ref{fig:layer_schematic}).  Radio emission peaks once the density ahead of the shock decreases sufficiently to reduce the free-free optical depth to a value of order unity.  The peak time and shape of the radio light curve thus depends on the density and radial profile of the DES.

\section{Dynamical Model}
\label{sec:model}

\subsection{Physical Set-Up}
\label{sec:setup}

\subsubsection{Dense External Shell}

The following radial profile is assumed for the density of the DES:
\begin{eqnarray}
n_4(r,t) = \frac{\chi M_{\rm DES}}{4\pi m_p \Delta r^{2}}\times\left\{
\begin{array}{lr}
 \exp\left[-(r-r_0)^{2}/\Delta^{2}\right], 
&(r < r_0) \\
\exp\left[\frac{k^{2}(r-r_0)}{k(r-r_0) + \Delta}\right]\left(\frac{k(r-r_0)}{\Delta}+1\right)^{-k},      
 &(r \ge r_0), 
\end{array}
\right.
\label{eq:n4}
\end{eqnarray}
where $r_0(t)$ and $\Delta(t)$ are the characteristic radial scale and thickness, respectively, of the DES at time $t$ (see below), external to which a power-law cut-off in radius is assumed, and $\chi$ is a constant chosen such that $\int 4\pi m_{\rm p} n_4 r^{2}dr$ equals the total mass of the DES, $M_{\rm DES}$.  The single peak in the mass distribution at $r \sim r_0$ implied by equation (\ref{eq:n4}) is justified by the requirement to produce the single observed maximum in the early radio, X-ray, and gamma-ray light curves of novae, while still providing flexibility regarding the key parameters (peak density, power-law index $k$) to which the emission is most sensitive.  The seemingly complex functional form of $n_4$ for $r \gtrsim r_0$ is chosen to enforce continuity of the derivative at $r = r_0$.

\begin{figure}
\includegraphics[width=0.5\textwidth]{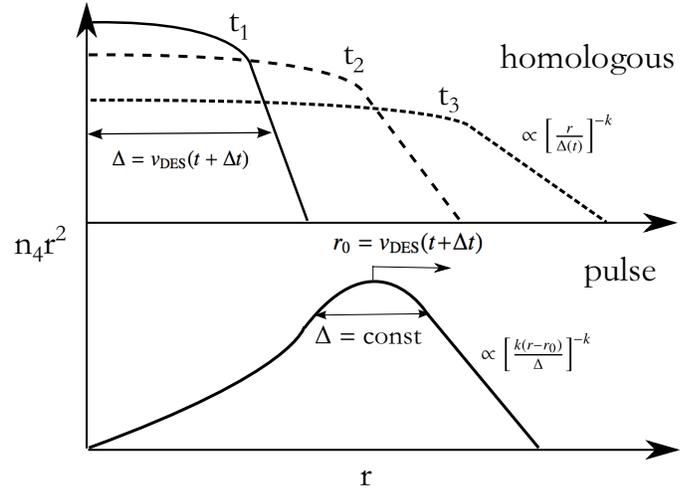}
\caption{Schematic diagram of the two models for the density structure $n_4(r,t)$ of the dense external shell (DES), both accommodated within the general form of equation (\ref{eq:n4}).  In the pulse model ({\it top panel}), the radial profile of the DES maintains a constant thickness $\Delta = v_{\rm DES}\Delta t$ about the characteristic radius $r_0 = v_{\rm DES}(t + \Delta t)$ expanding with velocity $v_{\rm DES}$.  In the homologous model ({\it bottom panel}), the density of the DES instead maintains a constant shape scaled to a characteristic width $\Delta = v_{\rm DES}(t + \Delta t)$ that instead grows with time at the characteristic velocity $v_{\rm DES}$.  The velocity of the DES is constant at all radii in the pulse model ($v_4 = v_{\rm DES}$), while the velocity varies with radius and time as $v_4 = r/(t + \Delta t)$ in the homologous model.} 
\label{fig:schematic_density}
\end{figure}

Two models are considered for the radial structure of the DES, both encompassed by the general form of equation (\ref{eq:n4}), as summarized in Figure \ref{fig:schematic_density}.  The first model assumes the DES is undergoing homologous expansion from the surface of the white dwarf, i.e.
\be 
r_0 = 0;\,\,\,\,\,\,\,\Delta = v_{\rm DES}(t + \Delta t),
\ee
where $v_{\rm DES} < v_{\rm w}$ is the (constant) characteristic velocity of the DES and $\Delta t > 0$ is the delay between the onset of the homologous expansion of the DES and the fast outflow (hereafter $t$ is defined with respect to the time at which the mass loss rate of the fast outflow peaks; see eq.~[\ref{eq:Mdot}] below).  The velocity of the DES in the homologous case varies with radius and time according to
\be
v_4 = r/(t + \Delta t),
\label{eq:homovel}
\ee
such that all fluid elements can be traced back to the origin at $t = -\Delta t$.  

The second model assumes that the DES is a shell of fixed width $\Delta$ with a constant velocity $v_4 = v_{\rm DES}$ at all radii (a `pulse'), in which case
\be
\Delta = v_{\rm DES}\Delta t = {\rm constant};\,\,\,\,\,\,\,r_0 = v_{\rm DES}(t + \Delta t),
\ee
where again $\Delta t$ represents the delay between the onset of the pulse and the peak of the fast outflow.  

Whether a homologous or pulse model is more physical depends on the origin of the DES.  If the DES represents a prior episode of mass loss from the white dwarf, then the `pulse' scenario is more appropriate assuming that the outflow possesses a well-defined characteristic velocity, such that the density structure maintains a coherent shape (memory of its initial physical scale $\Delta = v_{\rm DES}\Delta t$) at the larger radii of interest.  Homologous expansion will instead be approached if the velocity dispersion is sufficiently high to wipe out the initial spatial scale by the radii of interest, such that the outflow can be characterized as a point-like `explosion' from the origin.  If the DES represents ambient matter unrelated to the thermonuclear runaway, which of these scenario (if any) is more appropriate depends on what mechanism is responsible for accelerating the DES to its observed velocity.

The fast nova outflow of velocity $v_{\rm w}$ catches up to and collides with the slower DES of velocity $\sim v_{\rm DES}$ at a characteristic radius
\be
r_{\rm c} \sim \frac{v_{\rm DES} v_w\Delta t}{v_w - v_{\rm DES}} \sim 6\times 10^{13}\,\,v_{\rm w,8}\frac{v_{\rm DES}}{0.3v_w}\left(\frac{\Delta t}{20{\rm\, d}}\right)\,\,{\rm cm},
\label{eq:rcoll}
\ee
where in the second line we have assumed $v_{\rm DES} \lesssim v_w$ and $v_{w,8} \equiv v_{w}/10^{3}$ km s$^{-1}$.  The delay between the fast and slow outflows $\Delta t$ is normalized to a value $\sim 20$ d characteristic of the nova optical duration, which is assumed to coincide with the duration of significant mass loss. 

For a thick DES ($\Delta \sim r_{\rm c}$) with a steep outer density profile with $k > 1$ (as will be required to explain the fast evolution of the early radio light curves; $\S\ref{sec:results}$), the mass is concentrated at radii $r \sim r_{\rm c} \sim \Delta$.  The characteristic density $n_{\rm c}$ in equation (\ref{eq:n4}) at the time of collision can thus be estimated as
\be
n_{\rm c} \approx \frac{M_{\rm DES}}{4\pi m_p r_{\rm c}^{2}\Delta} \sim 10^{10}\left(\frac{M_{\rm DES}}{10^{-4}M_{\odot}}\right)\left(\frac{\Delta}{r_{\rm c}}\right)^{-1}\left(\frac{r_{\rm c}}{10^{14}{\rm cm}}\right)^{-3}{\rm cm^{-3}},
\label{eq:nin}
\ee
where $M_{\rm DES}$ is the total mass, normalized to a value $\sim 10^{-4}M_{\odot}$ characteristic of the total ejecta mass of nova (e.g.~\citealt{Seaquist&Bode08}).   Note that number density hereafter is defined as $n \equiv \rho/m_p$, where $\rho$ is the mass density.  

\subsubsection{Fast Nova Outflow}

The density profile of the fast nova outflow $n_1$ also depends on the nature of mass loss from the white dwarf.  If the outflow velocity $v_w$ varies slowly as compared to the flow time $\sim r/v_w$ at radii of interest, then the density of the nova outflow is well approximated by that of a steady wind, i.e.~
\be n_1 = \frac{\dot{M}_w}{4\pi v_w m_p r^{2}} \approx 3\times 10^{9}\dot{M}_{-3}v_{\rm w,8}^{-1}\left(\frac{r}{10^{14}{\rm cm}}\right)^{-2}{\,\rm cm^{-3}},
\label{eq:n1}
\ee 
where the nova mass loss rate $\dot{M}_{-3} = \dot{M}_w/10^{-3}M_{\odot}$ yr$^{-1}$ is normalized to a value characteristic of that required to eject $\sim 10^{-5}-10^{-4}M_{\odot}$ (e.g.~Roy et al.~2012) over a timescale of weeks to months characteristic of the optical duration of the nova.  A high initial value $\dot{M}_w \sim 10^{-3}M_{\odot}$ yr$^{-1}$ is also demanded by the large shock power $\sim \dot{M}_w v_w^{2}/2  \sim 10^{38}$ erg s$^{-1}$ required to explain the high observed gamma-ray luminosity (e.g.~$L_{\gamma} \gtrsim 3\times 10^{35}$ erg s$^{-1}$ for V1324 Sco; \citealt{Hill+13}) given a reasonable efficiency for accelerating relativistic particles.    

For simplicity we adopt the steady wind approximation $n_1 \propto \dot{M}_w/r^{2}$ (eq.~[\ref{eq:n1}]) for a fixed outflow velocity $v_{\rm w}$ and a mass loss rate $\dot{M}_{\rm w}$ with a simple time evolution
\be
\dot{M}_w = \dot{M}_{w,0}\left[\exp\left(\frac{-2\Delta  t_{\rm w}}{t+\Delta t_{\rm w}}\right)\right]\left(\frac{t+\Delta t_{\rm w}}{\Delta t_{\rm w}}\right)^{-2}.
\label{eq:Mdot}
\ee
The peak mass loss rate $\dot{M}_{w,0}$ is achieved at time $t = 0$ and lasts for a characteristic duration $\sim \Delta t_{\rm w}\sim$ weeks which is also assumed to be similar to the optical duration of the nova.  At late times $t \gg \Delta t_{\rm w}$, $\dot{M}_w$ decays as a power law $\propto t^{-2}$ similar to that predicted for the late optically-thick wind phase (\citealt{Bath&Shaviv76}; \citealt{Kato83}; \citealt{Kwok83}; \citealt{Prialnik86}; \citealt{Yaron+05}).  
Finally, note that even if the actual outflow is not well approximated as a steady wind, modeling of the late radio data of some nova finds that a $\propto 1/r^{2}$ density profile accurately characterizes the ejecta (e.g.~\citealt{Seaquist&Palimaka77}; \citealt{Weston+13}).  

The fast outflow and DES are assumed to be composed primarily of hydrogen, with a potentially high mass fraction of intermediate mass elements $X_Z \gtrsim 0.1$.  Such high ejecta metallicities $Z \gtrsim 10Z_{\odot}$ are inferred for the ejecta based on nova spectra (e.g.~\citealt{Livio&Truran94}) and are expected if the ejecta has mixed with the inner layers of the white dwarf.  If the DES originates from the binary companion instead of the white dwarf surface, then its metallicity would likely be substantially lower, $Z \sim Z_{\odot}$, than that of the ejecta.

\subsubsection{Shock Interaction}

Because the nova outflow velocity $v_{\it w}$ greatly exceeds the sound speed of the DES of temperature $T_4 \lesssim 10^{4}$ K, their interaction is mediated by shocks.  A forward shock of velocity $v_{\rm f}$ propagates into the DES, while a reverse shock of velocity $v_{\rm r}$ simultaneously propagates back into the nova outflow (\citealt{Lloyd+92}).  Gas behind the shocks cools, either rapidly if cooling is efficient and the shock is radiative, or on the longer expansion timescale if the gas cools adiabatically.  Gas collected by both shocks accumulates downstream in a cold ($\sim 10^{4}$ K) shell between the shocks.

In summary, the system is characterized by five physical regions, as shown in Figure \ref{fig:schematic}:
\begin{enumerate}[1.\hspace{0.3cm}]
\item{Unshocked nova ejecta of density $n_1$ (eq.~[\ref{eq:n1}]) and constant velocity $v_1 = v_w$.}
\item{Shocked ejecta of pressure $P_2$ and immediate post-shock density $n_2$, velocity $v_2$, and temperature $T_2 \simeq P_{2}/kn_2$.}
\item{Shocked DES of pressure $P_3$ and immediate post-shock density $n_3$, velocity $v_3$, and temperature $T_3 \simeq P_3/kn_3$.}
\item{Unshocked DES of density $n_4$ (eq.~[\ref{eq:n4}]) and velocity $v_4$, the latter of which is constant in the pulse model but varies with time and radius in the homologous model (eq.~[\ref{eq:homovel}]). }
\item{Cold shell separating regions 2 and 3 of temperature $T_{\rm sh} \approx 10^{4}$ K, mass $M_{\rm sh}$, radius $R_{\rm sh}$, and velocity $V_{\rm sh}$, the latter of which will in general differ from the immediate velocities of the gas behind the reverse and forward shocks ($v_2$ and $v_3$).}
\end{enumerate}
 Both the thickness of the central shell $\Delta_{\rm sh}$, as well as that of the post-shock region separating the shocks from the shell, are much less than the shell radius $R_{\rm sh}$ at times of interest; the latter can thus also be taken as the characteristic radius of both the forward and reverse shocks.

\subsection{Shock Jump Conditions}
\label{sec:jump}
The jump conditions at the forward shock are given by (\citealt{Landau&Lifshitz59})
\begin{eqnarray}
n_3 = 4 n_4; \\
n_3(v_3 - v_{\rm f}) = n_4(v_4 - v_{\rm f}); \\
P_3/m_p  = n_4(v_4 - v_{\rm f})^{2} - n_3(v_{3}-v_{\rm f})^{2},
\end{eqnarray}
while those at the reverse shock are given by
\begin{eqnarray}
n_2 = 4 n_1; \\
n_2(v_2 - v_{\rm r}) = n_1(v_w - v_{\rm r}); \\
P_2/m_p  = n_1(v_w-v_{\rm r})^{2} - n_2(v_{2}-v_{\rm r})^{2},
\end{eqnarray} 
from which it follows that
\be
v_{2} = \frac{1}{4}v_{\rm w} + \frac{3}{4}v_{\rm r};\,\,\,\,\,v_3 =  \frac{1}{4}v_{4} + \frac{3}{4}v_{\rm f};
\label{eq:vel}
\ee
\be
P_{2} = \frac{3}{4}m_p n_1 (v_{\rm w} - v_{\rm r})^{2};\,\,\,\,\,P_{3} = \frac{3}{4}m_p n_4 (v_{\rm f} - v_4)^{2},
\label{eq:pressure}
\ee
where again $v_{\rm f}$ and $v_{\rm r}$ are the velocity of the forward and reverse shocks, respectively.

The mass and momentum of the cold shell sandwiched between the shocks evolve according to
\be
\frac{dM_{\rm sh}}{dt} = \dot{M}_w\left(\frac{v_{\rm w}-v_{\rm r}}{v_{\rm w}}\right) + 4\pi R_{\rm sh}^{2}m_p n_4(v_{\rm f}-v_4);
\label{eq:dMshdt}
\ee
\be
\frac{d}{dt}\left(M_{\rm sh}v_{\rm sh}\right) = \dot{M}_w(v_{\rm w}-v_{\rm r})  + 4\pi R_{\rm sh}^{2}m_p n_4 v_4(v_{\rm f}-v_4),
\label{eq:dpshdt}
\ee
where the first(second) terms on the right hand side of equation (\ref{eq:dMshdt})[\ref{eq:dpshdt}] correspond to mass[momentum] swept up by the reverse(forward) shock.  

Solving equations (\ref{eq:vel})-(\ref{eq:dpshdt}) requires relating the properties of the flow immediately behind the shocks to those downstream at the interface of the central cold shell.  This relationship depends on cooling efficiency behind the shock, as characterized by the critical ratios,
\be
\eta_3 \equiv \frac{t_{\rm c,3}}{R_{\rm sh}/v_{\rm sh}};\,\,\,\,\,\,\,\,\,\,\,\,\eta_2 \equiv \frac{t_{\rm c,2}}{R_{\rm sh}/v_{\rm sh}},
\label{eq:eta}
\ee
where $t_{\rm c} \equiv P/n^{2}\Lambda$ is the cooling time of the post-shock gas, $R_{\rm sh}/v_{\rm sh}$ is the dynamical  timescale, and $\Lambda(T)$ is the temperature-dependent cooling function.  There are two limiting cases, depending on whether each shock is radiative ($\eta \ll 1$) or adiabatic ($\eta \gg 1$).

\subsubsection{Radiative Shocks ($\eta \ll 1$)}
Gas that cools efficiently behind the shock ($\eta \ll 1$) at constant pressure must slow to a velocity that matches that of both the shock and the central shell ($\S\ref{sec:coolinglayer}$), i.e.
\be
v_{\rm r} = v_{\rm sh}  \,\,\,\,\,\,\,(\eta_2 \ll 1);
\label{eq:vrrad}
\ee
\be
v_{\rm f} = v_{\rm sh}   \,\,\,\,\,\,\,(\eta_3 \ll 1).
\label{eq:vfrad}
\ee
The jump conditions (eq.~[\ref{eq:pressure}]) in this case give
\be
P_2 = \frac{3}{4}m_p n_1 (v_w - v_{\rm r})^{2} =  \frac{3}{4}m_p n_1 (v_w - v_{\rm sh})^{2}  \,\,\,\,\,\,\,(\eta_2 \ll 1);
\label{eq:P2rad}
\ee
\be
P_{3} = \frac{3}{4}m_p n_4 (v_{\rm f}-v_4)^{2} = \frac{3}{4}m_p n_4 (v_{\rm sh}-v_4)^{2}\,\,\,\,\,\,\,(\eta_3 \ll 1).
\label{eq:P3rad}
\ee
Note that in this case the velocities just behind the shock, $v_2$ and $v_3$, {\it do not} match those of the central shell (see eq.~[\ref{eq:vel}]).

\subsubsection{Adiabatic Shocks ($\eta \gg 1$)}
If cooling is inefficient ($\eta \gg 1$) then the velocity is approximately uniform in the post-shock region and equal to that of the central shell, i.e.~
\be 
v_{2} = v_{\rm sh}\,\,\,\,\,\,\,\,\,(\eta_2 \gg 1);
\ee
\be
v_3 = v_{\rm sh}\,\,\,\,\,\,\,\,\,\,(\eta_3 \gg 1),
\ee
such that (eq.~[\ref{eq:pressure}])
\be
P_{2} = \frac{3}{4}m_p n_1 (v_{w}-v_{\rm r})^{2} = \frac{4}{3}m_p n_1 \left(v_{\rm w} - v_{\rm sh}\right)^{2}\,\,\,\,\,\,\,\,\,(\eta_2 \gg 1);
\label{eq:P2ad}
\ee
\be
P_{3} = \frac{3}{4}m_p n_4 (v_{\rm f}-v_4)^{2} = \frac{4}{3}m_p n_4 (v_{\rm sh}-v_4)^{2}\,\,\,\,\,\,\,\,\,\,(\eta_3 \gg 1).
\label{eq:P3ad}
\ee

\subsection{Post Shock Regions}

\subsubsection{Mechanical Power}

Mechanical energy is converted to  thermal energy by the reverse and forward shocks at rates given, respectively, by
\begin{eqnarray}
\dot{E}_{\rm r} &\approx& \frac{4\pi R_{\rm sh}^{2}(v_w-v_{\rm r})}{4}\frac{3}{2}P_{2} = \frac{9}{32}\dot{M}\frac{(v_{\rm w} - v_{\rm r})^{3}}{v_{\rm w}};
\label{eq:edotr}
\end{eqnarray}

\begin{eqnarray}
\dot{E}_{\rm f} &\approx& \frac{4\pi R_{\rm sh}^{2}(v_{\rm f}-v_4)}{4}\frac{3}{2}P_{3} \nonumber \\
&\approx& 6\times 10^{38}n_{\rm c,10}v_{\rm w,8}^{3}\left(\frac{R_{\rm sh}}{10^{14}\,\rm cm}\right)^{2}\left(\frac{\tilde{v_{\rm f}}}{v_{\rm w}}\right)^{3}\frac{n_4}{n_{\rm c}}\,{\rm erg\,s^{-1}}, \nonumber \\
\label{eq:edotf}
\end{eqnarray}
where $P_2$ and $P_3$ are determined by equations [\ref{eq:P2rad},\ref{eq:P3rad},\ref{eq:P2ad},\ref{eq:P3ad}], the density of the DES is normalized to its characteristic value $n_{\rm c,10} \equiv n_{\rm c}/10^{10}$ g cm$^{-3}$ (eq.~[\ref{eq:nin}]) at the collision radius $r_{\rm c} \sim 10^{14}$ cm (eq.~[\ref{eq:rcoll}]), and we have defined $\tilde{v_{\rm f}} \equiv v_{\rm f} - v_4$ as the velocity of the forward shock in the rest frame of the upstream gas. 

If the shocks are radiative their bolometric luminosities can thus be comparable to those from the photosphere of the white dwarf itself, as the latter is approximately limited to the Eddington luminosity $\sim 10^{38}$ erg s$^{-1}$.  

\subsubsection{Post Shock Temperatures}
\label{sec:temp}

The temperatures just behind the reverse and forward shocks are given, respectively, by
\begin{eqnarray}
T_2 = \frac{P_2}{k n_2} = \frac{3m_p}{16k}(v_{\rm w}-v_{\rm r})^{2};
\label{eq:T2}
\end{eqnarray}
\begin{eqnarray}
T_3 &=& \frac{P_3}{k n_3} = \frac{3}{16}\frac{m_p}{k}\tilde{v_{\rm f}}^{2} 
\approx 2.2\times 10^{7}v_{\rm w,8}^{2}\left(\frac{\tilde{v_{\rm f}}}{v_{\rm w}}\right)^{2}{\rm K}. 
\label{eq:T3}
\end{eqnarray}
The post shock temperature is thus sensitive to the velocity structure of the outflow.  If the forward shock moves fast into the DES, i.e. $v_4 \ll v_{\rm f}$ ($\tilde{v_{\rm f}} \sim v_w$), then the shocked gas radiates at high temperatures $T \sim 10^{7}-10^{8}$ K corresponding to X-ray energies $kT \sim 1-10$ keV.  On the other hand, if $\tilde{v_{\rm f}} \ll v_{\rm w}$ then the characteristic temperature can be substantially lower, $\lesssim$ keV.  As discussed below in $\S\ref{sec:Xrays}$, much of this soft X-ray emission is initially absorbed by the largely neutral DES ahead of the shock (Region 4).  

\subsubsection{Cooling Layer}
\label{sec:coolinglayer}
We now consider the structure of the density $n$ and temperature $T$ behind the forward shock in greater detail (similar considerations apply to the reverse shock).  After being heated to a high temperature $T_3$ (eq.~[\ref{eq:T3}]), gas cools as it flows downstream away from the shock, such that its enthalpy evolves approximately according to   
\be
v_{\rm rel}\frac{d}{d z}\left(\frac{5}{2}kT\right) = -n\Lambda(T),
\label{eq:Tevo}
\ee
where $z$ is the distance behind the shock, $v_{\rm rel}$ is the velocity of the fluid in the frame of the shock, $\Lambda(T)$ is the cooling function, and we have adopted a one-dimensional model since the post shock layer is necessarily thin $\ll R_{\rm sh}$.  Note that equation (\ref{eq:Tevo}) neglects thermal conduction, an assumption that we justify in $\S\ref{sec:conduction}$.  The velocity is determined by constant mass flux, 
\be
n v_{\rm rel} = n_3(v_{\rm f}-v_{3}) =  n_4(v_{\rm f}-v_4),
\label{eq:constflux}
\ee for the steady downstream flow.  The shocked gas cools at constant pressure $P \propto nT = n_3 T_3$, such that
\be
n = n_3(T_3/T).
\label{eq:constP}
\ee
Cooling at constant pressure is justified because the velocity $v_{\rm rel} \propto T$ decreases faster than the sound speed $c_{\rm s} \propto T^{1/2}$ downstream; sound waves from the post shock region thus always remain in causal contact with the cooling gas on a timescale short compared to the evolution timescale.  Since the post shock gas may cool to temperatures as low as $\sim 10^{4}$ K in a fairly thin layer, its density can increase by a factor up to $\sim 10^{2}-10^{4}$ (assuming that magnetic and cosmic ray pressure can be neglected; $\S\ref{sec:discussion}$).  When the shock is radiative, the post shock gas thus slows (in the frame of the shock) to a velocity essentially matching that of the shock, as we have assumed in deriving equations (\ref{eq:vrrad}) and (\ref{eq:vfrad}).

At high temperatures ($T \gtrsim 10^{6}$ K), cooling is dominated by free-free emission, while at lower temperatures ($T \lesssim 10^{6}$ K) line cooling may instead dominate.  In $\S\ref{sec:ion}$ we show that UV/X-ray radiation from the shock keeps the post shock gas over-ionized relative to its state in collisional ionization equilibrium, which may suppress line cooling even at lower temperatures where it would normally dominate free-free cooling.

If free-free emission dominates the cooling rate $\Lambda = \Lambda_{\rm ff} \propto T^{1/2}$ (eq.~[\ref{eq:lambdaff}]), then equation (\ref{eq:Tevo}) can be solved using equations (\ref{eq:constflux}) and (\ref{eq:constP}) to obtain
\be
T(z) = T_3\left(1 - \frac{z}{L_{\rm cool}}\right)^{2/5},
\label{eq:Tff}
\ee
where $L_{\rm cool}$ is the cooling length given by
\begin{eqnarray}
\frac{L_{\rm cool}}{R_{\rm sh}} \equiv \frac{t_{\rm c,3}}{R_{\rm sh}/(v_{\rm f}-v_{3})} \approx 2\times 10^{-3}v_{\rm w,8}^{2}n_{\rm c,10}^{-1}\left(\frac{\tilde{v_{\rm f}}}{v_{\rm w}}\right)^{2}\left(\frac{n_4}{n_{\rm c}}\frac{R_{\rm sh}}{10^{14}\rm cm}\right)^{-1},
\label{eq:tcool2}
\end{eqnarray}
the ratio of the initial cooling timescale of the post-shock gas $t_{\rm c,3} \equiv P_{3}/n_{3}^{2}\Lambda_{\rm ff}|_{T_{3}}$ to the downstream flow time $\sim R_{\rm sh}/(v_{\rm f}-v_3)$, and
\be
\Lambda_{\rm ff} \simeq \Lambda_0 T^{1/2} = 2\times 10^{-27}T^{1/2}\sum\limits_{i} \frac{Z_i^{2}X_{i}}{A_i}\sum\limits_{j} \frac{Z_j X_{j}}{A_j} {\rm erg\,cm^{3}\,s^{-1}}
\label{eq:lambdaff}
\ee 
is the free-free cooling function (e.g.~\citealt{Rybicki&Lightman79}), where the Gaunt factor has been absorbed into the prefactor.  In using equation (\ref{eq:lambdaff}), we have also assumed fully ionized gas and that X-rays freely escape the post-shock region at times of interest; such {\it effectively} optically thin cooling will be justified in $\S\ref{sec:postshock}$.  The sums in $\Lambda_{\rm ff}$ extend over all atomic species of charge $Z_i$, mass $A_i$ and mass fraction $X_i$, with their product varying only from unity by a modest factor $\lesssim 2$ for compositions ranging from relatively low metallicity $Z \lesssim 10Z_{\odot}$ to one completely dominated by intermediate mass elements (we hereafter assume the value of this factor to be unity).    

The prediction of equation (\ref{eq:Tff}) that the temperature goes to zero a distance $L_{\rm cool}$ behind the shock is an artificial outcome of our assumptions.  The rate of free-free cooling was calculated assuming complete ionization, whereas in reality the ejecta will recombine at sufficiently low temperatures.  Line cooling, which requires an excited population of atoms, also becomes ineffective at low temperature.  Although recombination will thus limit the post shock gas to a minimum temperature $\sim 10^{4}$ K, equation (\ref{eq:Tff}) nevertheless provides a reasonable first approximation of the profile above this minimum temperature.

Note that because $\eta_3$ increases $\propto 1/n_4$ the shock is radiative at early times, before evolving to become adiabatic at late times as the density decreases.  An expression similar to equation (\ref{eq:tcool2}) applies also to the temperature structure behind the reverse shock (Region 2), except for the difference that $L_{\rm cool}$ is defined in terms of the velocity $v_{2} - v_{\rm r}$ (eq.~[\ref{eq:eta}]). 

When cooling behind the forward shock is inefficient ($\eta_3 \gg 1$) only a fraction $\Delta T/T_{3} \sim 2/5\eta_3 \ll 1$ of the thermal energy produced at the shock is radiated within a distance $\ll R_{\rm sh}$ of the shock, with the rest lost to adiabatic expansion (a similar argument applies to the reverse shock).  A useful approximate expression for the radiated luminosity of the forward and reverse shocks, which smoothly connects the radiative ($\eta \ll 1$) and adiabatic ($\eta \gg 1$) limits, is thus given by
\be
L_{\rm r} \simeq \frac{\dot{E}_{\rm r}}{1 + 5\eta_2/2};\,\,\,\,\,\,\,\,\,\,\,\,\,L_{\rm f} \simeq \frac{\dot{E}_{\rm f}}{1 + 5\eta_3/2}, 
\label{eq:Lf}
\ee
where $\dot{E}_{\rm r}$ (eq.~[\ref{eq:edotr}]) and $\dot{E}_{\rm f}$ (eq.~[\ref{eq:edotf}]) are the total mechanical powers dissipated by the shocks.

\subsubsection{Thermal Conduction}
\label{sec:conduction}

In deriving equation (\ref{eq:Tff}) we have neglected the effects of thermal conduction on the temperature profile behind the shock.  Conduction adds a term to the right hand side of equation (\ref{eq:Tevo}) of the form 
\be
\nabla \cdot \vec{F_{\rm c}} \simeq  \frac{d}{dz}\left(\kappa \frac{dT}{dz}\right),
\label{eq:conduction}
\ee
where $\vec{F_{\rm c}}$ is the conductive flux, $\kappa = \kappa_0 T^{5/2}$ is the conductivity, $\kappa_0 \approx 2\times 10^{-5}/{\rm ln\Lambda}$  is an (approximate) constant in cgs units, and ln$\Lambda \approx 10$ is the Coulomb logarithm (\citealt{Spitzer62}).  The importance of conduction on the structure of the post shock cooling layer depends on the ratio of the conductive heating to radiative cooling rates:
\begin{eqnarray}
\frac{\nabla \cdot \vec{F_{\rm c}}}{n^{2}\Lambda} \sim \frac{\kappa_0 T^{7/2}/(\eta_3 R_{\rm sh})^{2}}{n^{2}\Lambda_0 T^{1/2}} \underset{T \approx T_3}\approx 0.08\left(\frac{T_{3}}{10^{7}{\rm K}}\right)\left(\frac{\rm ln\Lambda}{10}\right)^{-1},
\label{eq:condratio}
\end{eqnarray}
which we have approximated by substituting solution (\ref{eq:Tff}) into equation (\ref{eq:conduction}).  The fact that this ratio is small for typical values of the forward shock temperature $T_3 \sim 10^{7}$ K implies that we may safely neglect the effects of thermal conductivity hereafter.


\subsection{Ionization State}
\label{sec:ion}

The ionization state of the DES and the post-shock cooling layer are of great relevance to calculating the observed X-ray and radio emission from the forward shock.

\subsubsection{Unshocked DES}

In the absence of external sources of photo-ionization, the unshocked DES (Region 4) will be largely neutral in regions where the density is high, as is justified by the short timescale for recombination $t_{\rm rec} \sim 1/n_{4}\alpha_{\rm rec}^{Z}$,
\be
\frac{t_{\rm rec}}{r/v_{w}} \sim 1\times 10^{-4}v_{\rm w,8}n_{\rm c,10}^{-1}\left(\frac{n_{4}}{n_{\rm c}}\right)^{-1}\left(\frac{r}{10^{14}\rm cm}\right)^{-1}
\ee
compared to the minimum dynamical time $\sim r/v_{w}$, where $\alpha_{\rm rec}^{Z} \approx 1.1\times 10^{-12}Z^{2}(T/10^{4}{\rm K})^{-0.8}$ cm$^{3}$ s$^{-1}$ is the approximate radiative recombination rate for hydrogen-like atomic species of charge $Z$ (\citealt{Osterbrock&Ferland06}), which in the above numerical estimate is evaluated for hydrogen gas ($Z = 1$) of temperature $T_4 = 10^{4}$ K.

The DES is subject to ionizing UV and X-ray radiation from the forward shock.  Similar radiation from the reverse shock, or from the surface of the white dwarf, is instead likely to be absorbed by the central cold neutral shell before reaching the DES (Fig.~\ref{fig:schematic}).  Radiation from the forward shock penetrates the DES to a depth $\Delta^{Z}$ (see eq.~[\ref{eq:deltaion}] below) that depends on the element $Z$ which dominates the bound-free opacity at the frequency of relevance.  The neutral fraction of this exposed layer of width $\Delta^{Z}$,
\be f_{\rm n,Z} = (1 + \lambda_{Z})^{-1},
\ee
is set by the balance between ionization and recombination, where
\begin{eqnarray}
\lambda_{Z} &\equiv& \frac{4\pi}{\alpha_{\rm rec}^{Z}n_e}\int \frac{J_{\nu}}{h\nu}\sigma_{\rm bf}^{Z}(\nu)d\nu \simeq \frac{L_{\rm f}\sigma_{\rm thr}^{Z}}{12\pi \alpha_{\rm rec}^{Z} n_4 R_{\rm sh}^{2}kT_{3}}\left(\frac{J_{\nu}}{J_{\rm f}}\right) \nonumber \\
&\approx& \frac{\sigma_{\rm thr}^{Z}\tilde{v_{\rm f}}}{2\alpha_{\rm rec}^{Z}(1 + 5\eta_3/2)}\left(\frac{J_{\nu}}{J_{\rm f}}\right) \nonumber \\
&\sim&  \frac{3\times 10^{2}v_{\rm w,8}}{(1 + 5\eta_{3}/2)}Z^{-4}\left(\frac{T_4}{10^{4}\rm K}\right)^{0.8}\left(\frac{\tilde{v_{\rm f}}}{v_{\rm w}}\right)\left(\frac{J_{\nu}}{J_{\rm f}}\right)
\label{eq:fn}
\end{eqnarray}
is the ratio of the rates of photo-ionization and recombination.  Here $n_e \sim n_4$ is the electron density\footnote{Approximating the gas as being fully ionized is justified here because at all times of interest hydrogen, which supplies most of the electrons, is almost completely ionized (neutral fraction $f_{\rm n,Z = 1} \ll 1$) in the ionized layers of higher $Z$ elements.}, $\sigma_{\rm bf}^{Z} \simeq \sigma_{\rm thr}^{Z}(\nu/\nu_{\rm thr}^{Z})^{-3}$ is the (approximate) bound-free cross section, and $\sigma_{\rm thr}^{Z} \approx 6\times 10^{-18}Z^{-2}$ cm$^{2}$ is the (approximate) cross section at the ionization threshold frequency $h\nu_{\rm thr}^{Z} \approx 13.6Z^{2}$ eV (\citealt{Osterbrock&Ferland06}).  

In equation (\ref{eq:fn}), $J_{\nu}$ is the intensity of ionizing photons normalized to that, $J_{\rm f} \approx L_{\rm f}h/16\pi^{2}R_{\rm sh}^{2}kT_3$, from the forward shock of luminosity $L_{\rm f}$ (eq.~[\ref{eq:Lf}]), which we assume is produced by free-free cooling with its characteristic flat spectrum at frequencies $h\nu \ll kT_3$ (eq.~[\ref{eq:T3}]).  This direct luminosity from the shock sets the {\it minimum} ionizing intensity, but $J_{\nu}$ can exceed $J_{\rm f}$ if higher frequency radiation is absorbed by the neutral gas and then reprocessed to the lower frequency band of relevance.  Unless otherwise noted, hereafter we assume $J_{\nu} = J_{\rm f}$.  This is because our primary application of equation (\ref{eq:fn}) is to calculate: (1) the luminosity of $\sim 1$ keV X-rays; by the time the DES is transparent to $\sim 1$ keV photons, even harder radiation is also free to escape and hence will not be re-processed to lower energies; (2) The depth of the hydrogen ionization layer at late times relevant to the radio peak.  At such times the DES is also likely to be transparent to the EUV radiation ($h\nu \gtrsim 13.6$ eV) of greatest relevance to hydrogen ionization.  

Equation (\ref{eq:fn}) shows that when the forward shock is radiative at early times ($\eta_3 \ll 1$) we have $f_{\rm n,Z} \ll 1$ for $Z \lesssim 5$.  Lighter elements in the layer of the DES directly exposed to radiation from the shock are thus completely ionized.  However, at later times as the radiative efficiency of the shock decreases ($\eta_3$ increases), the layer of DES directly exposed to the forward shock becomes more neutral $f_{\rm n,Z} \propto 1/\eta_{3}$ .  For intermediate mass elements $Z \gtrsim 5$, corresponding to X-ray energies $\gtrsim$ keV, the `ionized layer' is in fact almost fully neutral ($\lambda_Z \ll 1$; $f_{\rm n,Z} \approx 1$) at all times.  

Radiation of frequency $\nu \sim \nu_{\rm thr}^{Z}$ penetrates the neutral gas to the depth $\Delta^{Z}$ where the absorptive optical depth equals unity, i.e.~ 
\be (n_4/A)(\sigma_{\rm thr}^{Z}/2)X_Z f_{\rm n,Z}\Delta^{Z} \sim  1,
\ee
where $X_Z$ is the mass fraction of element $Z$ and we have taken a characteristic cross section $\sigma_{\rm thr}^{Z}/2$ as half that of the threshold value (see Appendix A of \citealt{Metzger+14} for justification).  For $\Delta^{Z} \ll R_{\rm sh}$ we thus have (in the $\lambda_Z \gg 1$ and $\lambda_Z \ll 1$ limits)
\begin{eqnarray}
\frac{\Delta^{Z}}{R_{\rm sh}} &\underset{\lambda_Z \gg 1}\approx&  \frac{A}{X_{Z}(1 + 5\eta_{3}/2)}\frac{\tilde{v_{\rm f}}}{n_4 R_{\rm sh}\alpha_{\rm rec}^{Z}}\left(\frac{J_{\nu}}{J_{\rm f}}\right) \nonumber \\
&\sim&  \frac{9\times 10^{-5}}{1+5\eta_{3}/2}\frac{v_{\rm w,8}}{X_Z}\frac{A}{Z^{2}}n_{\rm c,10}^{-1}\left(\frac{\tilde{v_{\rm f}}}{v_{\rm w}}\right)\left(\frac{n_4}{n_{\rm c}}\frac{R_{\rm sh}}{10^{14}\rm cm}\right)^{-1}\left(\frac{J_{\nu}}{J_{\rm f}}\right) \\
\frac{\Delta^{Z}}{R_{\rm sh}} &\underset{\lambda_Z \ll 1}\approx& \frac{2A}{n_4 \sigma_{\rm thr}^{Z}X_{Z}R_{\rm sh}} \sim 3\times 10^{-7}\frac{AZ^{2}}{X_{Z}}n_{\rm c,10}^{-1}\left(\frac{n_4}{n_{\rm c}}\frac{R_{\rm sh}}{10^{14}\rm cm}\right)^{-1}
\label{eq:deltaion}
\end{eqnarray}
where we have used equation (\ref{eq:fn}) and have assumed $T_4 = 10^{4}$ K for the temperature of the photo-ionized layer ahead of the shock.  The ionized layer separating the forward shock from the neutral DES is thus initially thin, but its width relative to the shock radius expands with time, usually as $\Delta^{Z} \propto n_{4}^{-1}$, before saturating to a constant value once the shock is no longer radiative.  Ionizing radiation (UV or X-rays) is free to pass through the DES, and hence to reach the external observer, once $\Delta^{Z}$ exceeds the shell thickness ($R_{\rm sh}$ or $\Delta$, depending on the model for the DES) ($\S\ref{sec:Xrays}$).  For $\lambda_{Z} \ll 1$ elements with higher $Z$ are ionized to a larger depth $\Delta^{Z} \propto A Z^{2}/X_Z$, so hard X-rays are free to escape first.  

\subsubsection{Post Shock Cooling Layer}
\label{sec:postshock}

The rate $\Lambda$ at which gas cools behind the forward shock depends on the ionization state of the post shock gas.  Another important issue is thus whether ionizing photons from the shock can penetrate the cooling layer downstream of the forward shock.  The bound-free optical depth through the cooling layer to a depth $z$ (or, equivalently, down to a temperature $T$) behind the shock is given by
\begin{eqnarray}
&&\tau_{\rm bf,3}^{Z}(T) = \int_0^{z} n \frac{\sigma_{\rm thr}^{Z}X_{Z} }{2A}f_{\rm n,Z}(z')dz' \simeq \frac{\sigma_{\rm thr}^{Z}X_{Z} n_3}{2A}\int_{T_3}^{T} \frac{T_3}{T'}f_{n,Z}\left(\frac{dT'}{dz'}\right)^{-1}dT' \nonumber \\
&&\approx \frac{5\sigma_{\rm thr}^{Z} k X_{Z}\tilde{v_{\rm f}}}{16A T_{3}}\int_{T_{3}}^{T} \frac{T'}{\Lambda(T')}\frac{1}{\lambda_{Z}}dT' \approx \frac{5k X_Z}{2A\Lambda_0}\left(\frac{J_{\rm f}}{J_{\nu}}\right)\int_{T_{3}}^{T}\frac{\alpha_{\rm rec}^{Z}}{(T')^{1/2}}dT' \nonumber \\
&&\sim 5.8 \left(\frac{X_Z}{0.1}\right)\frac{Z^{2}}{A}\left(\frac{J_{\rm f}}{J_{\nu}}\right)\left(\frac{T_{3}}{10^{7}{\rm K}}\right)^{-0.1}\left(\left(\frac{T}{T_3}\right)^{-0.1}-1\right),
\label{eq:taubf3}
\end{eqnarray}
where in the top line we have used equation (\ref{eq:constP}) and calculate the temperature gradient using equation (\ref{eq:Tevo}).  In the second line the neutral fraction 
\be
f_{\rm n,Z} \approx \frac{1}{\lambda_Z} \approx 0.7\left(\frac{Z}{6}\right)^{4}v_{w,8}^{-1}\left(\frac{T_{3}}{10^{7}{\rm K}}\right)\left(\frac{\tilde{v_{\rm f}}}{v_{\rm w}}\right)^{-1}\left(\frac{T}{10^{6}{\rm K}}\right)^{-1.6}\left(\frac{10 J_{\rm f}}{J_{\nu}}\right)
\label{eq:fn2}
\ee
is calculated using equation (\ref{eq:fn}) in the fast cooling case $\eta_3 \ll 1$ of relevance but replacing the density of the DES, $n_4$, with the density $n > n_3$ of the post shock cooling layer.  The final equality in equation (\ref{eq:taubf3}) makes use of the numerical approximation $\alpha_{\rm rec}^{Z} \sim 4\times 10^{-14}Z^{2}(T/10^{6}{\rm K})^{-0.6}$ cm$^{3}$ s$^{-1}$ for the recombination rate, as is a reasonable approximation for hydrogen-like ionization states and temperatures $\sim 10^{6}$ K.

Equation (\ref{eq:taubf3}) shows that the optical $\tau_{\rm bf,3}^{Z}$ is of order unity or less for UV photons with energies corresponding to light elements such as He ($X_{\rm He} \sim 0.1; Z = 2; A = 4$) down to temperatures $T \simeq 10^{6}$ K of relevance to setting the radio emission (eq.~[\ref{eq:Tobspeak}]), assuming a typical value for the forward shock temperature $T_3 = 10^{6.5}-10^{7}$K (Fig.~\ref{fig:various1}).  A similar conclusion applies to soft X-rays corresponding to intermediate light elements such as C ($X_{\rm C} \sim 0.03; Z = 6; A = 12$).  We conclude that ionizing photons may penetrate the post-shock layer to depths and temperatures relevant to the radio emission.

That ionizing radiation may penetrate the cooling layer is relevant because commonly used cooling functions (e.g.~\citealt{Schure+09}) assume collisional ionization equilibrium (CIE), which is accurate only if photo-ionization can be neglected.  Over-ionized plasmas, such as those present during the late radiative stages of supernova remnants (e.g.~\citealt{Kafatos73}; \citealt{Sutherland&Dopita93}) and in intergalactic space (\citealt{Wiersma+09}), are known to cool at rates lower than those of a plasma in collisional ionization equilibrium.  This is because cooling due to bound-bound transitions is suppressed at a given temperature if the bound atomic levels are under-populated due to photo-ionization.  Because line cooling is dominated by CNO elements, we are primarily interested in the ionization states of these elements.    

Whether line cooling is in fact suppressed requires both that ionizing radiation penetrate the post shock layer and that it result in a much lower neutral fraction than that predicted by CIE.  Equation (\ref{eq:fn2}) suggests that hydrogen-like states of CNO elements are unlikely to possess a neutral fraction much below unity if the ionizing radiation field is just the direct radiation from the forward shock, i.e. $J_{\nu} = J_{\rm f}$.  However, if the UV/soft X-ray flux is greatly enhanced ($J_{\nu} \gtrsim 10 J_{\rm f}$) due to the re-processing of higher energy X-rays absorbed by the DES or the post-shock plasma (see discussion following eq.~[\ref{eq:fn}]) then $f_{\rm n,Z}$ could indeed be sufficiently low ($\ll 1$) to greatly reduce the plasma cooling rate.  

Although equation (\ref{eq:fn2}) provides a useful estimate of the neutral fraction of hydrogen-like atomic species, a low neutral fraction according to this criterion may be an overly strict requirement to suppress line cooling if the latter is dominated by bound-bound transitions of lower ionization states.  On the other hand, the recombination rates$-$and hence neutral fraction$-$of such states can also be significantly higher than naively estimated by equation (\ref{eq:fn2}) for hydrogen-like atoms.  A more detailed analysis of the ionization structure behind the post-shock layer, including a better treatment of the radiation transport of the forward shock emission, would be necessary to assess whether photo-ionization can indeed suppress line cooling sufficiently to justify our assumption that free-free cooling dominates at temperatures $T \sim 10^{6}$ K.  


\section{Observed Emission}
\label{sec:emission}

With the dynamics and ionization state of the system described, we now move on to discuss the emission from nova shock interaction.  Different physical processes are responsible for X-ray, optical and radio emission.

\subsection{X-rays}
\label{sec:Xrays}

The high predicted peak luminosities of nova shocks ($L_{\rm r}, L_{\rm f} \sim 10^{37}-10^{38}$ erg s$^{-1}$ for characteristic parameters; eq.~[\ref{eq:Lf}]) would at first appear to be easily detectable at typical distances of Galactic novae $\sim$ few kpc.  Relatively hard X-ray emission is indeed observed in coincidence with some novae (e.g.~\citealt{Ness+07}; \citealt{Mukai+08}), but the observed luminosities $L_{\rm X} \lesssim 10^{35}$ erg s$^{-1}$ are typically much lower than those predicted at early times by equation (\ref{eq:Lf}).  Furthermore, in other cases where shock interaction is clearly present, no X-ray emission is detected to deep upper limits (e.g.~$L_X \lesssim 10^{33}$ erg s$^{-1}$ for V1324 Sco; $\S\ref{sec:novasco}$).  

This apparent contradiction with observations is explained by the fact that X-ray emission can be delayed or suppressed due to absorption by the DES ahead of the forward shock.  The {\it observed} luminosity from the forward shock at frequency $\nu_X$ is given by
\be
L_{\rm X} \equiv L_{\nu_{\rm X}}\nu_{\rm X} = L_{\rm f}\frac{h\nu_{\rm X}}{kT_{3}}\exp\left[\frac{-h\nu_{\rm X}}{kT_{3}}\right]\exp(-\tau_{\rm \nu_{\rm X},4}),
\label{eq:Lx}
\ee
where $L_{\rm f}$ (eq.~\ref{eq:Lf}) is the luminosity of the forward shock and 
\begin{eqnarray}
\tau_{\rm \nu_{\rm X},4} &\simeq& \int_{R_{\rm sh}}^{\infty} \frac{\sigma_{\rm thr}^{Z}}{2}\frac{X_{Z} n_4}{A}f_{\rm n,Z}dr \nonumber \\
&\approx& 2.5\times 10^{2}\left(\frac{X_Z}{0.1}\right)\left(\frac{h\nu_{\rm X}}{\rm keV}\right)^{-3/2}n_{\rm c,10}r_{\rm c,14}\frac{\int n_{4}dr}{n_{\rm c}r_{\rm c}}
\label{eq:taubf}
\end{eqnarray}
is the bound-free optical depth of the unshocked DES, where in the numerical estimate we assume that the dominant opacity at frequency $\nu_{\rm X}$ originates from element $Z$ of ionization potential $h\nu_X \sim h\nu_{\rm thr}^{Z} \sim 0.5(Z/6)^{2}{\rm keV}$.  We also assume $A = 2Z$ and $f_{\rm n} = 1$, in accordance with our conclusion from equation (\ref{eq:fn}) for intermediate mass elements (e.g.~carbon, $Z = 6$) of relevance. 

Equation (\ref{eq:taubf}) shows that the optical depth to $\sim$ 1 keV X-rays greatly exceeds unity at early times, but that $\tau_{\rm X,4}$ decreases with time as the radius of the shock increases and hence the absorbing column decreases (the latter evolution being consistent with X-ray observations of novae; e.g.~\citealt{Mukai&Ishida01}).  The X-ray luminosity at $\nu_{\rm X}$ peaks once the optical depth $\tau_{\rm X,4}$ decreases to $\lesssim 1$.  The time at which this optically thin transition occurs, and hence the resulting peak X-ray luminosity, is sensitive to the abundance of heavy CNO nuclei $X_Z$ and to the temperature of the forward shock $T_3$ (eq.~[\ref{eq:T3}]), the latter of which is itself a sensitive function of the velocity $\tilde{v_{\rm f}}$ of the forward shock with respect to the DES.  The net result of this sensitivity is that the X-ray luminosity may vary considerably depending on the properties of the nova outflow and the DES.

Even though the luminosity of the reverse shock often exceeds that of the forward shock, the reverse shock X-ray emission is probably much weaker due to the additional absorption by the cold neutral central shell (Fig.~\ref{fig:schematic}).  In what follows we assume that the X-ray luminosity originates entirely from the forward shock according to equation (\ref{eq:Lx}).

\begin{figure*}
\includegraphics[width=0.8\textwidth]{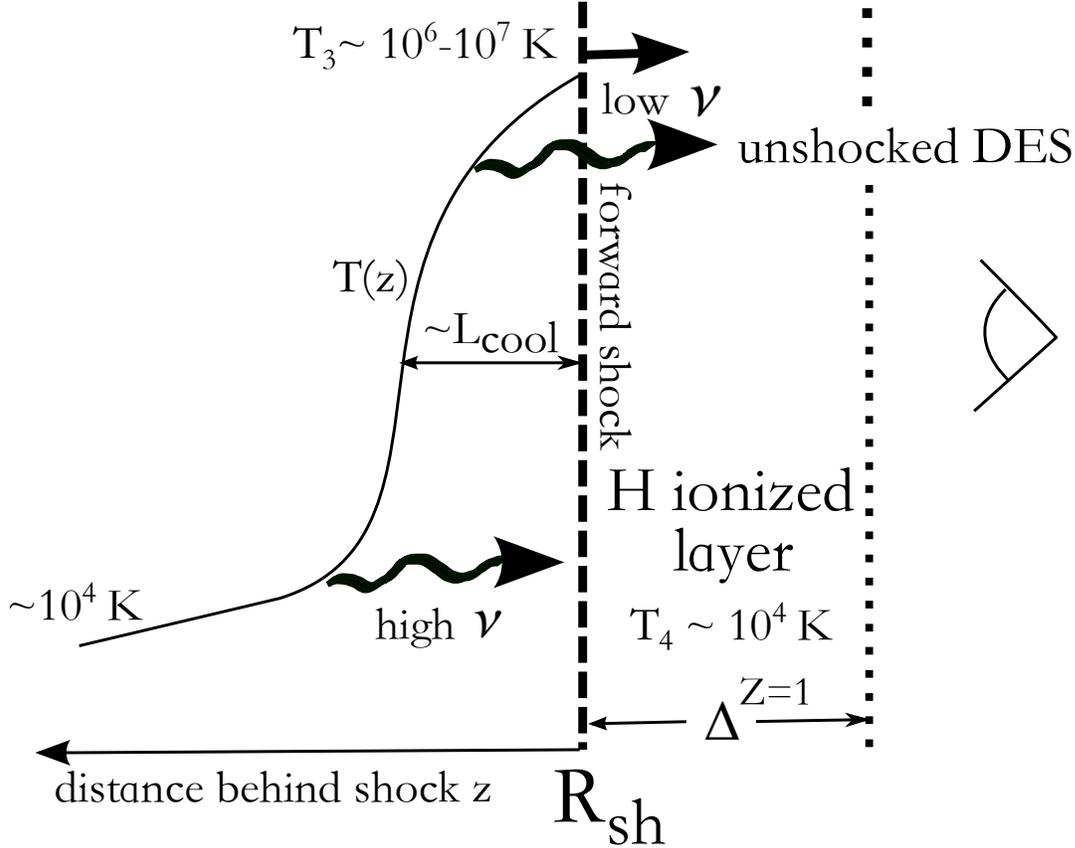}
\caption{Schematic diagram of radio emitting region from the forward shock when it is radiative ($\eta_3 \ll 1$) at early times.  Gas rapidly cools within a short distance $\sim L_{\rm cool}$ (eq.~[\ref{eq:Tff}]) behind the shock, producing a dense layer.  Radio emission originates from the free-free photosphere at temperature $T_{\nu}$, the value of which is frequency dependent.  The unshocked DES ahead of the forward shock (Region 4) is ionized by UV/X-rays from the shock out to a distance $\Delta^{Z = 1} \ll R_{\rm sh}$ (eq.~[\ref{eq:deltaion}]).  The radio brightness temperature measured by an external observer, $T_{\nu}^{\rm obs}$ is initially lower than $T_{\nu}$ due to free-free absorption by this ionized layer.  The effect of this frequency-dependent absorption is to cancel the frequency dependence of brightness temperature at the shock $T_{\nu}$, rendering the {\it measured} brightness temperature at the time of peak flux $T_{\nu}^{\rm peak}$ (eq.~[\ref{eq:Tobspeak}]) to be frequency independent.  The light curve at higher frequencies also peaks earlier in time.} 
\label{fig:layer_schematic}
\end{figure*}

\subsection{Optical}
\label{sec:optical}

Hard UV and X-ray photons absorbed by the DES are re-radiated at lower energies.  This emission will freely escape once it is degraded to optical radiation ($h\nu \ll 13.6$ eV) where the optical depth
\be
\tau_{\rm opt,4} \simeq \int_{R_{\rm sh}}^{\infty} m_p n_4\kappa_{\rm opt} dr \approx 0.7\left(\frac{\kappa_{\rm opt}}{\kappa_{\rm es}}\right)n_{\rm c,10}r_{\rm c,14}\frac{\int n_{4}dr}{n_{\rm c}r_{\rm c}},
\label{eq:tauopt4}
\ee
is much lower than at hard UV/X-ray frequencies (eq.~[\ref{eq:taubf}]).  In the numerical estimate the optical opacity $\kappa_{\rm opt}$ is scaled to the electron scattering opacity $\kappa_{\rm es} \simeq 0.4$ cm$^{2}$ g$^{-1}$ as a characteristic value, although other higher sources of opacity, such as Doppler-broadened line absorption, are more appropriate in certain frequency ranges.  The timescale required for an optical photon to escape the DES, $t_{\rm diff} \sim (\tau_{\rm opt,4}+1)(R_{\rm sh}/c)$, is much shorter than the expansion time of the shock through the upstream matter $\sim \Delta^{Z}/(v_{\rm f}-v_4)$.  This allows the re-processed radiation to escape before being trapped behind the dense post-shock gas and hence degraded by adiabatic expansion.

To the extent that a fraction $\epsilon_{\rm opt}$ of the X-ray luminosities of the shocks (eq.~[\ref{eq:Lx}]) absorbed by the DES is re-radiated as an optical/UV luminosity, the latter is approximately given by
\be
L_{\rm opt} = \epsilon_{\rm opt}L_{\rm f}[1 - \exp(-\tau_{\rm \nu_{\rm X} = kT_{3},4})] + \epsilon_{\rm opt}L_{\rm r},
\label{eq:Lopt}
\ee
where $\tau_{\rm \nu_{\rm X} = kT_{3},4}$ is the optical depth of X-rays with frequency $h\nu = kT_{3}$ corresponding to the temperature $T_3$ behind the forward shock, and we have assumed that X-rays from the reverse shock are absorbed by the central cool shell at all times.  The optical efficiency $\epsilon_{\rm opt}$ is highly uncertain because it depends in part on what fraction of the energy absorbed as hard X-rays is able to escape at intermediate wavelengths (e.g. hard UV or soft X-rays) before reaching the optical band.  In what follows we assume for simplicity a value $\epsilon_{\rm opt} = 0.1$ that is independent of time.  

The potentially large luminosity of the forward shock $L_{\rm opt} \sim L_{\rm f} \sim 10^{37}-10^{38}$ erg s$^{-1}$ at early times when $\tau_{\rm \nu_{\rm X},4} \gg 1$ suggests that shock-powered emission could contribute significantly to the optical light curves of novae ($\S\ref{sec:discussion}$), which are normally interpreted as being powered entirely by thermal energy generated by nuclear burning on the white dwarf surface (e.g.~\citealt{Hauschildt+97}).  In the above we have assumed that the reverse and forward shocks re-radiate a similar fraction $\epsilon_{\rm opt}$ of the absorbed shock luminosity as optical radiation.  An important caveat, however, is the possibility that dust may form in the dense central shell; dust formation would substantially increase $\kappa_{\rm opt}$ and hence reduce the contribution of the reverse shock to the optical emission (which would instead presumably emerge in the infrared).    

\subsection{Radio}
\label{sec:radio}

Gas heated by the forward shock also produces thermal radio emission of intensity $I_{\nu} = 2 kT_{\nu}\nu^{2}/c^{2}$, which can be calculated given the known temperature (eq.~[$\ref{eq:Tff}$]) and density (eq.~[$\ref{eq:constP}$]) profiles behind the shock.   Here $T_{\nu}$ is the brightness temperature of radio emission at frequency $\nu$, which is a useful quantity because it evolves with depth $z$ behind the shock in a simple manner as determined by the equation of radiative transfer (\citealt{Rybicki&Lightman79}) 
\be
\frac{dT_{\nu}}{d\tau_{\rm ff,\nu}} = T - T_{\nu},
\label{eq:TBevo}
\ee
where $T$ is the gas temperature, $\tau_{\rm ff,\nu} = \int\alpha_{\rm ff,\nu}dz$ is the free-free optical depth, and
\be
\alpha_{\rm ff, \nu} = \alpha_{0,\nu}T^{-3/2}n^{2} \approx 0.06 T^{-3/2}n^{2}\nu^{-2}\sum\limits_{i} \frac{Z_i^{2}X_{i}}{A_i}\sum\limits_{j} \frac{Z_j X_{j}}{A_j} {\rm cm^{-1}},
\label{eq:alphaff}
\ee
is the free-free absorption coefficient (\citealt{Rybicki&Lightman79}), where again the Gaunt factor is absorbed into the prefactor and the sum varies only modestly from $\sim 1-2$ depending on the composition of the gas (we hereafter assume a value of unity; see discussion after eq.~[\ref{eq:lambdaff}]).

There are two regimes of radio emission, depending on whether the forward shock is radiative or adiabatic ($\S\ref{sec:jump}$).  

\subsubsection{Radiative Forward Shock ($\eta_3 \ll 1$)}

\begin{figure}
\includegraphics[width=0.5\textwidth]{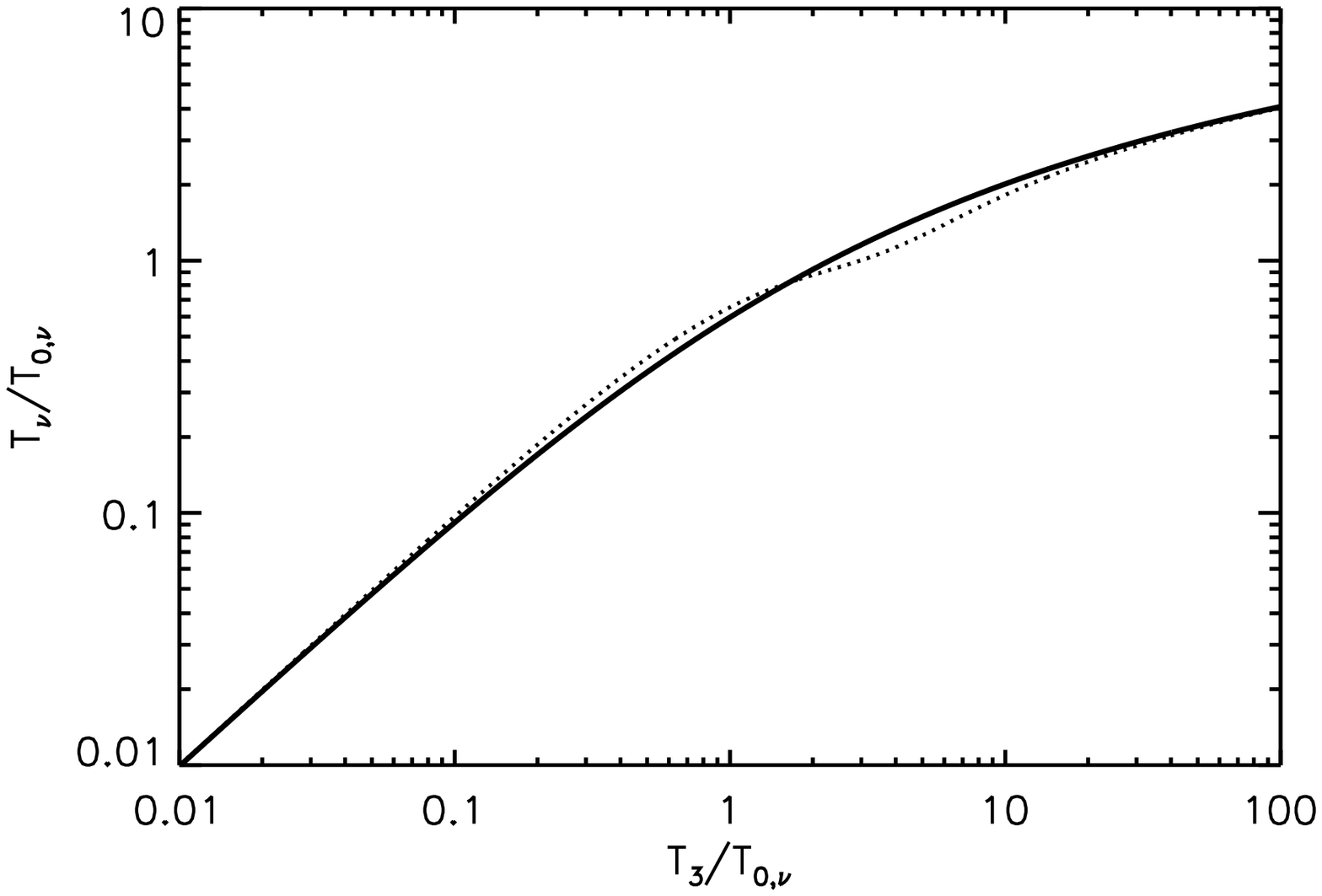}
\caption{Solution to equation (\ref{eq:eq2}) for the brightness temperature $T_{\nu}$ of emission as a function of immediate post shock temperature $T_3$ of the radiative shock.  The exact solution  (eq.~[\ref{eq:fullsol}]) is shown with a solid line, while the approximate solution (eq.~[\ref{eq:TBrad}]), accurate to $\lesssim 10\%$ at all temperatures, is shown by a dashed line.  Temperatures are normalized to the characteristic value $T_{\rm 0,\nu}$ (eq.~[\ref{eq:T0nu}]).  Note that $T_{\nu}$ must be corrected for free-free absorption by the unshocked material ahead of the shock to obtain the {\it observed} brightness temperature ($\S\ref{sec:obs}$).} 
\label{fig:layer}
\end{figure}

First consider the case when the shock is radiative ($\eta_3 \ll 1$; eq.~[\ref{eq:tcool2}]).  The hottest gas immediately behind the shock is optically thin to free-free emission ($\alpha_{\rm ff} \propto T^{-3/2}$; eq.~[\ref{eq:alphaff}]), but as gas cools downstream and becomes denser its optical depth increases rapidly.  Using the relationship
\be
d\tau_{\nu} = \alpha_{\rm ff,\nu}dz = \alpha_{\rm ff,\nu}(dT/dz)^{-1}dT = \frac{5P_3\tilde{v_{\rm f}}}{8}\frac{\alpha_{\rm ff,\nu}}{n^{2}\Lambda}\frac{dT}{T_3},
\ee
where $dT/dz$ follows from equation (\ref{eq:Tevo}) and the assumption of steady flow behind the shock (eq.~[\ref{eq:constflux}]), equation (\ref{eq:TBevo}) becomes
\be
\frac{dT_{\nu}}{dT} = \frac{5 k n_3\tilde{v_{\rm f}}}{8}\frac{\alpha_{\rm ff,\nu}}{n^{2}\Lambda}\left(T - T_{\nu}\right)
\ee
Using equation (\ref{eq:alphaff}) and again assuming free-free cooling $\Lambda = \Lambda_0 T^{1/2}$ (eq.~[\ref{eq:lambdaff}]), the above can be written as
\be
\frac{dT_{\nu}}{dT} = \frac{T_{0,\nu}}{T}\left(1 - \frac{T_{\nu}}{T}\right),
\label{eq:eq2}
\ee
where 
\be
T_{0,\nu} \equiv \frac{5}{8}\frac{\alpha_{0,\nu} k n_3\tilde{v_{\rm f}}}{\Lambda_0} \approx 1.0\times 10^{8}\nu_{10}^{-2}n_{\rm c,10}v_{\rm w,8}\left(\frac{n_4}{n_{\rm c}}\right)\left(\frac{\tilde{v_{\rm f}}}{v_{\rm w}}\right)\,{\rm K}.
\label{eq:T0nu}
\ee
For the desired boundary condition of LTE at high optical depth ($T_{\nu} = T$ at $T \ll T_{0,\nu}$), the solution to equation (\ref{eq:eq2}) for the brightness temperature of radiation emerging from the forward shock is given by
\be
T_{\nu} = T_{0,\nu}e^{1/x}E_{1}(1/x),
\label{eq:fullsol}
\ee
where $x \equiv T_{3}/T_{0,\nu}$ and $E_{1} \equiv \int_x^{\infty}u^{-1}e^{-u}du$ is the exponential integral (solid line in Fig.~\ref{fig:layer}).  Although $E_1$ cannot be expressed in terms of elementary functions, the following approximate solution for $T_{\nu}$ is accurate to $\lesssim 10\%$ across the entire range of temperature:
\begin{eqnarray}
T_{\nu} \approx T_{0,\nu}\left\{\left[{\rm ln}\left(x + 1\right) - \gamma\right]\left(1 - e^{-x}\right) + xe^{-x}(1+\gamma)\right\},
\label{eq:TBrad}
\end{eqnarray}
where $\gamma \simeq 0.577$ is the Euler constant (dashed line in Fig.~\ref{fig:layer}).  

For $T_3 \gg T_{0,\nu}$ (as is usually satisfied) the brightness temperature from the shock is thus $T_{\nu} \sim T_{0,\nu}$, up to a logarithmic correction of order unity.   This characteristic temperature $T_{0,\nu}$ can be interpreted as that of the `radio photosphere' (as observed at the forward shock), with the $\propto \nu^{-2}$ dependence indicating that higher frequency radiation originates from further behind the shock where the temperature is lower and the density is higher.  

Logarithmic corrections to $T_{\nu}$ can be understood physically by the fact that the intensity $I_{\nu} \propto T_{\nu}$ is proportional to the integrated free-free emissivity $\int j_{\nu}^{\rm ff}dz$ from the shock to the photosphere.  Using the facts that (1) $j_{\nu}^{\rm ff} \propto n^{2}T^{-1/2}$ for free-free emission; (2) $n \propto T^{-1}$ for cooling at constant pressure (eq.~[\ref{eq:constP}]);  and (3) $dT/dz \propto T^{-3/2}$ if the post shock temperature profile is itself set by by free-free cooling (eq.~[\ref{eq:Tevo}]), one concludes that $T_{\nu} \propto \int j_{\nu}^{\rm ff}dz \propto \int dT/T$ receives equal contributions from each decade of temperature behind the shock.

If the cooling rate is higher than the free-free rate we have assumed, this will act to lower the value of $T_{0,\nu} \propto \Lambda^{-1}$ but will not alter the $\nu^{-2}$ dependence as long as the dominant cooling mechanism $\Lambda \propto T^{\chi}$ has a temperature dependence $\chi \sim 1/2$.  Also note that the $T_{\nu} \propto \nu^{-2}$ dependence implies that the flux originating from behind the forward shock $F_{\nu} \propto T_{\nu}\nu^{2} \propto \nu^{0}$ is approximately independent of frequency (although this flat spectrum will not in general be observed since the radiation must first pass through the ionized screen provided by the unshocked DES; see $\S\ref{sec:obs}$ below).  This is a coincidence resulting from the frequency dependence of the temperature of the (optically thick) photosphere, largely unrelated to the similarly flat spectrum predicted for {\it optically thin} free-free emission.

Figure \ref{fig:layer_schematic} is a schematic figure of the post shock layer in the fast cooling case and its relationship to the radio emission.

\subsubsection{Adiabatic Forward Shock ($\eta_3 \gg 1$)}

When the shock is adiabatic ($\eta_3 \gg 1$; eq.~[\ref{eq:tcool2}]) the post-shock regions instead remains hot and optically thin behind the shock.  If the external density (and hence the post shock pressure) decreases sufficiently slowly with radius, then the radiation at a given time is dominated by matter swept by the shock within the most recent dynamical time.  An approximate solution to equation (\ref{eq:TBevo}) for the brightness temperature leaving the forward shock in this case is given by
\begin{eqnarray}
&T_{\nu}& \approx T_3\tau_{\rm ff,3} \nonumber \\
&\underset{\eta_3 \gg 1}\approx& 5\times 10^{9}\nu_{10}^{-2}v_{\rm w,8}^{-1}n_{\rm c,10}^{2}\left(\frac{\tilde{v_{\rm f}}}{v_{\rm w}}\right)^{-1}\left(\frac{n_{4}}{n_{\rm c}}\right)^{2}\left(\frac{R_{\rm sh}}{10^{14}{\rm cm}}\right){\rm K},
\label{eq:TBad}
\end{eqnarray}
where $\nu_{10} = \nu/$10 GHz and
\begin{eqnarray}
\tau_{\rm ff,3} = \alpha_{\rm ff,\nu}|_{T_{3},n_{3}}\frac{R_{\rm sh}}{4}  \simeq 220\nu_{10}^{-2}v_{\rm w,8}^{-3}n_{\rm c,10}^{2}\left(\frac{\tilde{v_{\rm f}}}{v_{\rm w}}\right)^{-3}\left(\frac{R_{\rm sh}}{10^{14}{\rm cm}}\right)\left(\frac{n_4}{n_{\rm c}}\right)^{2}
\end{eqnarray}
is the optical depth behind the shell, which is calculated under the approximation that the density and temperature are uniform a distance $\sim R_{\rm sh}$ behind the shock.  The high prefactor in equation (\ref{eq:TBad}) is deceptive, as this expression is valid only at late times when the shock is adiabatic and $\tau_{\rm ff,3} \lesssim 1$, i.e. once $n_4 \ll n_{\rm c}$ and the value of $T_{\nu}$ is substantially lower.

Equation (\ref{eq:TBad}) may not be a good approximation if the external density decreases sufficiently rapidly that the post shock gas cannot respond to the decrease in pressure behind the forward shock on the dynamical timescale.  Determining the radio emission in this case would require following the density/temperature structure of the adiabatically expanding gas, thus requiring a full hydrodynamical simulation.  Hereafter we adopt equation (\ref{eq:TBad}) in the adiabatic case, with the understanding that it probably represents a lower limit to the true brightness temperature because it does not account for the emission from gas swept up at earlier epochs.  Our results are fortunately insensitive to this assumption because the early radio emission peaks while the forward shock is still radiative.

\subsubsection{Observed Radio Emission}
\label{sec:obs}

Although equations (\ref{eq:TBrad}) and (\ref{eq:TBad}) represent the brightness temperature just in front of the forward shock, at early times this is not the temperature observed by a distant observer due to additional free-free absorption by the external DES.  The temperature measured by an external observer is given by
\be
T_{\nu}^{\rm obs} = T_{\nu}\exp(-\tau_{\rm ff, 4}) + T_{4}\left[1 - \exp(-\tau_{\rm ff,4})\right],
\ee
where $T_{4} \approx 10^{4}$ K is the temperature of the ionized unshocked DES (Region 4) and
\begin{eqnarray}
&\tau_{\rm ff,4}& = \int_{R_{\rm sh}}^{\infty} \alpha_{\nu}^{\rm ff}dr \approx \alpha_{0,\nu}T_{4}^{-3/2}n_{4}^{2}\Delta^{Z=1} 
\label{eq:tauff4}
\end{eqnarray}
is the free-free optical depth.  Here the effective thickness of the absorbing region is taken to be $\Delta^{Z = 1} \ll R_{\rm sh}$ (eq.~[\ref{eq:deltaion}]) since the free-free opacity depends only on the density of free electrons and ions, the greatest column of which is supplied by ionized hydrogen (the hydrogen-ionized region also extends to a greater distance through the DES than higher $Z$ elements). 

Radio emission peaks when $\tau_{\rm ff,4} \lesssim 1$, which from equation (\ref{eq:tauff4}) occurs when the shock reaches a DES density $n_4 \lesssim \alpha_{\rm rec}^{Z=1}|_{T_4}T_{4}^{3/2}(1 + 5\eta_3/2)/\alpha_{0,\nu}\tilde{v_{\rm f}}$.  Usually this optically thin transition occurs when the shock is still radiative ($\eta_3 \ll 1$), in which case this condition can be combined with equation (\ref{eq:TBrad}) to obtain an estimate of the peak observed brightness temperature:
\be
T_{\nu}^{\rm peak} \approx \frac{5 k T_{4}^{3/2}\alpha_{\rm rec}^{Z = 1}|_{T_4}}{2\Lambda_0}{\rm ln}\left[\frac{T_{3}}{T_{0,\nu}}\right] \approx 2\times 10^{5}{\rm ln}\left[\frac{T_{3}}{T_{0,\nu}}\right]{\rm K},
\label{eq:Tobspeak}
\ee
where $T_4 = 10^{4}$ K is again assumed in the numerical estimates.  The value of $T_{\nu}^{\rm peak} \sim 10^{6}$ K is typically a factor $\sim 100$ higher than the minimum temperature of the post shock gas $\sim 10^{4}$ K, but a factor $\sim 10$ times lower than the peak temperature just behind the forward shock $T_3 \sim 10^{7}$ K.   The relatively low value of $T_{\nu}^{\rm peak}$ at first appears inconsistent with our assumption that free-free emission dominates line cooling in the post-shock layer down to the photosphere.  However, as discussed in $\S\ref{sec:postshock}$, line cooling may be effectively suppressed if hard photons from the forward shock penetrate the post-shock region and over-ionize atomic species relevant to the line cooling.  

Equation (\ref{eq:Tobspeak}) predicts a peak temperature which is in practice {\it independent of frequency}, even though at any time the brightness temperature immediately ahead of the forward shock $T_{\nu} \propto \nu^{-2}$ (eqs.~[\ref{eq:TBad}, \ref{eq:TBrad}]) depends sensitively on frequency.  These statements are reconciled by the fact that lower frequencies peak later in time, at which point the density immediately behind the forward shock (to which the unshielded photosphere temperature is also proportional; eq.~[\ref{eq:T0nu}]) has decreased in a compensatory way.


The radio flux density observed by a distant observer is given by
\be
F_{\nu} = \frac{2\pi \nu^{2}R_{\rm sh}^{2}}{c^{2}}\frac{kT_{\nu}^{\rm obs}}{D^{2}},
\label{eq:Fnu}
\ee
where $D$ is the nova distance.

\section{Time Dependent Model}

\label{sec:results}

The results from $\S\ref{sec:model}$ and $\ref{sec:emission}$ are now brought together to calculate the time evolution of nova shocks and their associated radiation.

\subsection{Description and Numerical Procedure}
Equations (\ref{eq:dMshdt}) and (\ref{eq:dpshdt}) are solved for the evolution of the mass and momentum of the central shell.  The shock velocities and post-shock pressures are calculated at each time-step as described in $\S\ref{sec:jump}$, depending on whether the shocks are radiative ($\eta \ll 1$) or adiabatic ($\eta \gg 1$) using the definitions (\ref{eq:eta}).  The radius of the shell, which is also assumed to match that of both shocks, is calculated from the velocity as $dR_{\rm sh}/dt = V_{\rm sh}$.  

The densities of the DES and the fast nova outflow are calculated at each time $t$ and radius $R_{\rm sh}$ from equations (\ref{eq:n4}) and (\ref{eq:n1}), respectively, with the latter calculated from the mass loss evolution $\dot{M}_w(t)$ given in equation (\ref{eq:Mdot}).  When comparing to observations, the fast outflow is assumed to peak at the same time as the optical maxima, while the delay $\Delta t$ between the ejection of the DES and the fast peak (optical maximum) is assumed to correspond to the observed delay between the initial optical rise (the `outburst') and the optical peak (which for V1324 Sco was $\Delta t = 18$ days).  Finally, when employing the `pulse' model for the DES, the pulse width is also assumed to be set by the delay until the fast outflow, i.e. $\Delta = v_{\rm DES}\Delta t$.    

Once the evolution of the shell is determined, all dynamical quantities are smoothly interpolated between the radiative ($\eta \ll 1$) and adiabatic ($\eta \gg 1$) limits before radiation from the forward shock is calculated.  The 1 keV X-ray and optical luminosities are calculated from equation (\ref{eq:Lx}) and (\ref{eq:Lopt}), respectively, using the forward shock luminosity from equation (\ref{eq:Lf}).  The radio flux density is calculated from equation (\ref{eq:Fnu}) using the brightness temperature calculated according to equations (\ref{eq:TBad}), (\ref{eq:TBrad}).  Photo-ionized gas ahead of the shock is assumed to have a temperature $T_4 = 10^{4}$ K, with the same minimum temperature of $10^{4}$ K being imposed on the post shock gas (as motivated by the much lower cooling efficiency at lower temperatures).

\subsection{Results}

Our results are summarized in Figures \ref{fig:various1}-\ref{fig:radiolums} and Table \ref{table:results}.  We begin by describing our fiducial model in $\S\ref{sec:novasco}$, which has been chosen to provide a reasonable (though by no means unique) fit to the multi-wavelength observations of V1324 Sco.  Then in $\S\ref{sec:range}$ we describe our results for a wider diversity of models.  

\subsubsection{Example Case: V1324 Sco}
\label{sec:novasco}

\begin{figure}
\includegraphics[width=0.5\textwidth]{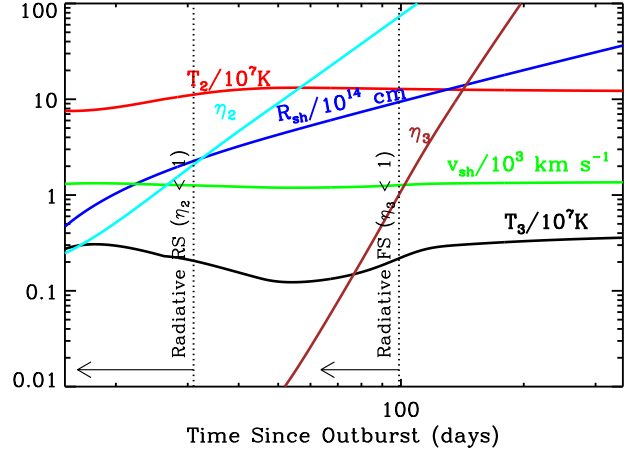}
\caption{Quantities relevant to the shock dynamics as a function of time (as measured from the onset of the optical outburst on June 2, 2012), calculated for our fiducial model corresponding to fast outflow velocity $v_w  = 3\times 10^{3}$ km s$^{-1}$; peak mass loss rate $\dot{M}_{w,0} = 4\times 10^{-4}M_{\odot}$ yr$^{-1}$; and mass loss duration $\Delta t_{\rm w} = 10$ days (eq.~[\ref{eq:Mdot}]).  The density of the DES is characterized by a pulse profile (Fig.~\ref{fig:schematic_density}) with a total total mass $M_{\rm DES} = 2\times 10^{-4}M_{\odot}$, characteristic velocity $v_{\rm DES} \approx 1000$ km s$^{-1}$, and density profile index k = 4.  Quantities shown include the radius $R_{\rm sh}$ ({\it blue}) and velocity $V_{\rm sh}$ ({\it green}) of the cold central shell; the temperatures $T_2$ ({\it red}) and $T_3$ ({\it black}) behind the reverse and forward shocks, respectively; and ratio $\eta$ of the cooling timescale to the flow time behind the forward ({\it fuscia}) and reverse ({\it cyan}) shocks.  The reverse and forward shocks transition from being radiative to being adiabatic as marked by the vertical dashed lines.}
\label{fig:various1}
\end{figure}

\begin{figure}
\includegraphics[width=0.5\textwidth]{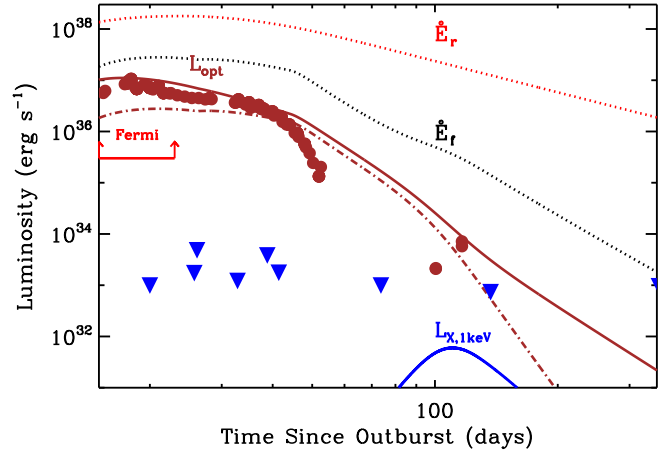}
\caption{Power dissipated by the reverse and forward shocks, $\dot{E}_{\rm r}$ ({\it red dashed}; eq.~[\ref{eq:edotr}]) and $\dot{E}_{\rm f}$  ({\it black dashed}; eq.~[\ref{eq:edotf}]), respectively, for the solution shown in Figure \ref{fig:various1}.  Also shown are the predicted 1 keV X-ray and optical/UV luminosities of the forward and reverse shocks, $L_X$ ({\it blue solid}; eq.~[\ref{eq:Lx}]) and $L_{\rm opt}$ ({\it fuscia solid}; eq.~[\ref{eq:Lopt}]).  The contribution of the optical luminosity originating just from the forward shock is shown as a dot-dashed line.  Shown for comparison is the gamma-ray luminosity of V1324 Sco as measured by {\it Fermi} (\citealt{Hill+13}), which can be interpreted as a strict lower limit on the total shock power $\dot{E}_r + \dot{E}_f$.  
Also shown are the {\it Swift} X-ray upper limits (\citealt{Page+12}; also see text) and the V-band optical light curve from the AAVSO (\citealt{Wagner+12}).  The UV light curve as measured by UVOT on {\it Swift} (not shown) peaked at $\sim 15-20$ days after the optical maximum (\citealt{Page+12}).  }
\label{fig:lum1}
\end{figure}

\begin{figure}
\centering
\subfigure{
\includegraphics[width = 0.48\textwidth]{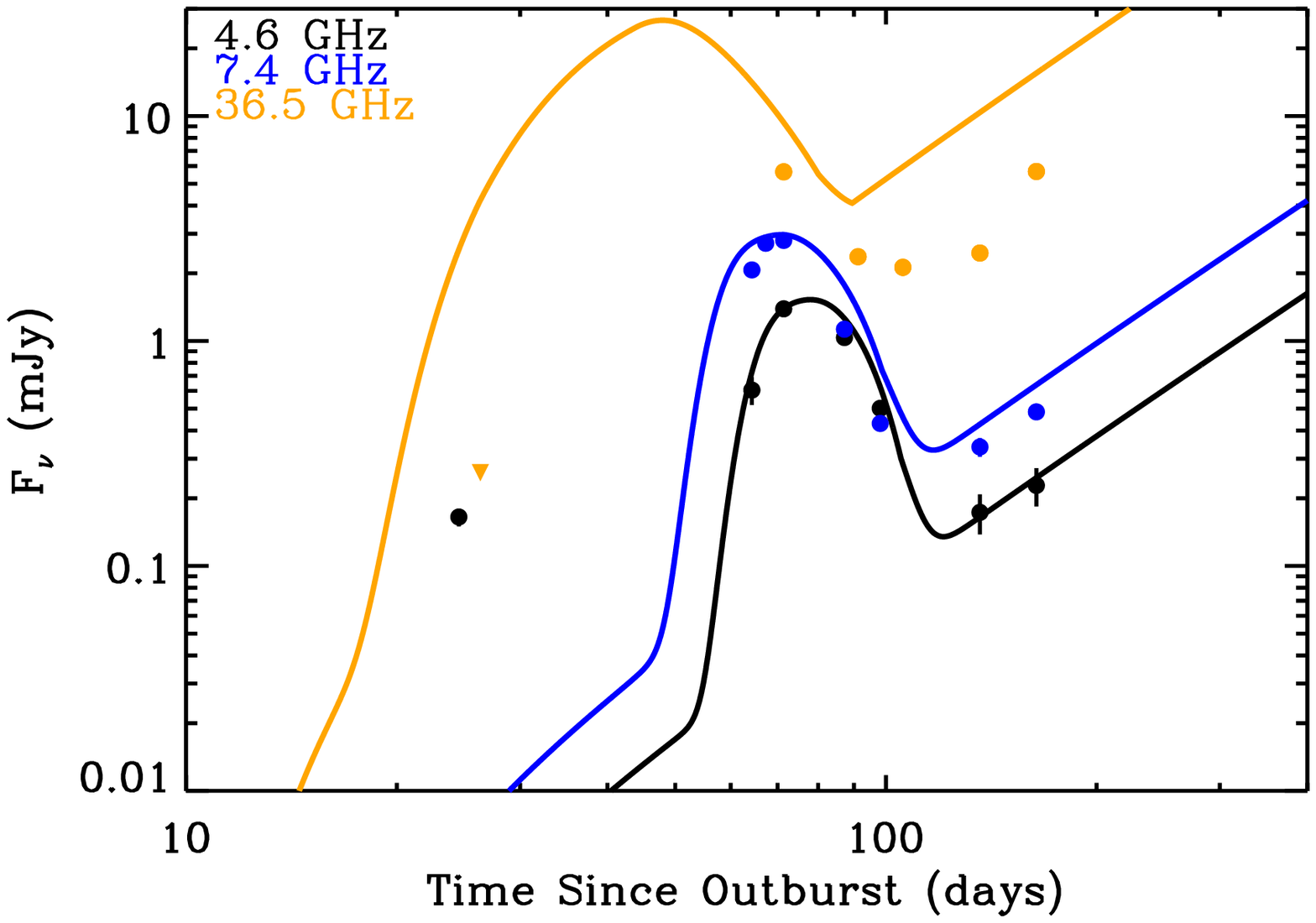}}
\subfigure{
\includegraphics[width = 0.48\textwidth]{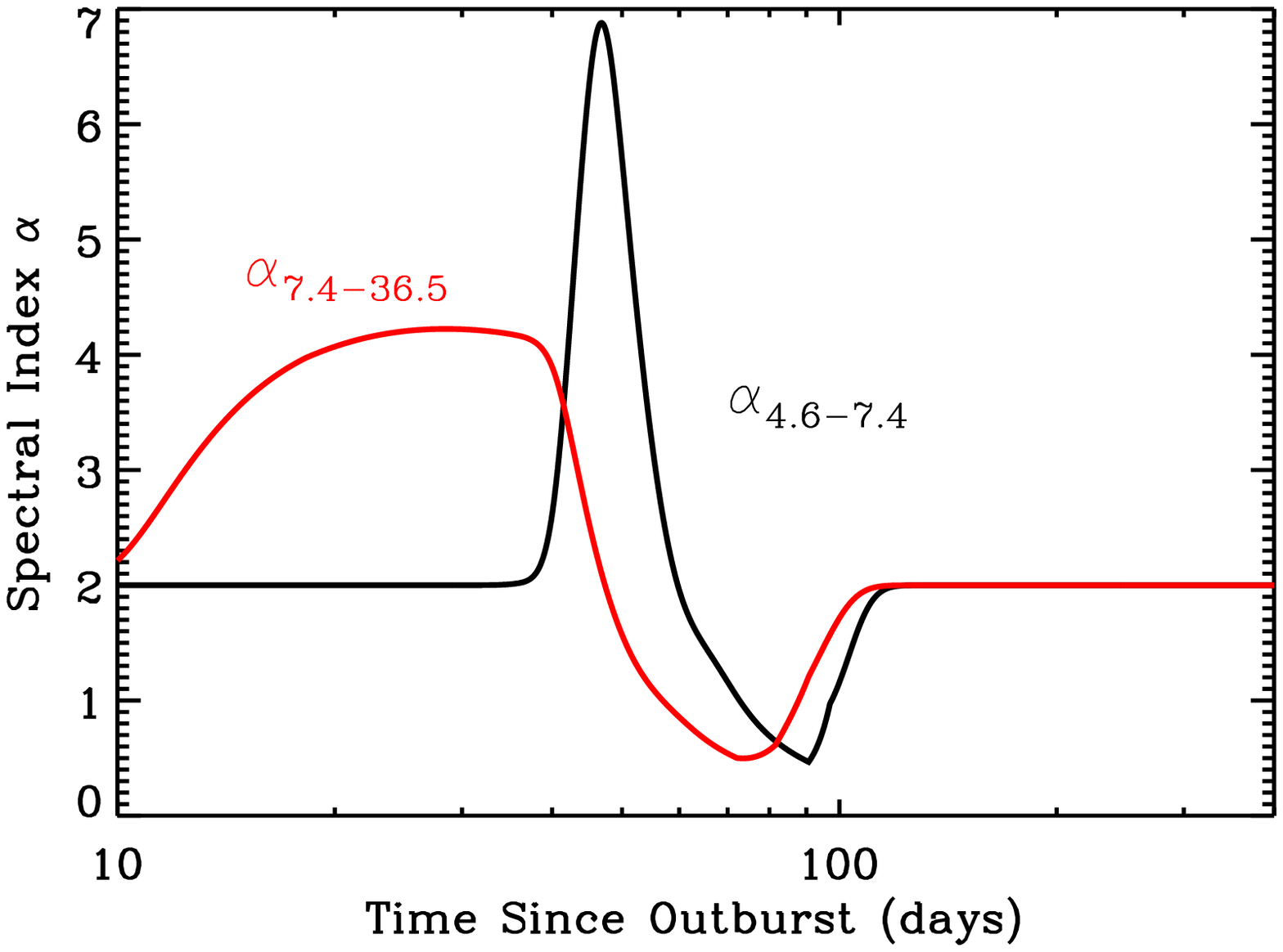}}
\caption[]{{\it \normalsize Top Panel:}  Radio luminosities for the same solution shown in Figures \ref{fig:lum1} and \ref{fig:various1} at frequencies $\nu = 4.6, 7.4, 36.5$ GHz (detections shown as circles; upper limit as a triangle).  Shown for comparison are observations of V1324 Sco from \citet{Finzell+14}, in prep. {\it \normalsize Bottom Panel:} Spectral index $\alpha$ as a function of time for the light curves shown in the top panel.  Phases at both early and late times with $\alpha = 2$ correspond to epochs when the emitting gas is optically thick with $T_4 \sim 10^{4}$ K.}
\label{fig:radio1}
\end{figure}

\begin{figure}
\includegraphics[width=0.5\textwidth]{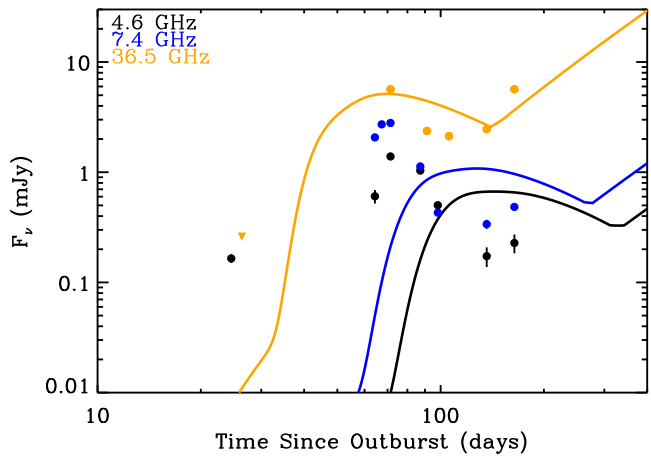}
\caption{Predicted radio luminosities versus multi-frequency observations of V1324 for a model with the same parameters as those in Figure \ref{fig:radio1}, but calculated for a homologously expanding DES with the same characteristic expansion velocity and the same outer density profile index $k = 4$.  The sharp radio peak achieved in the case of the pulse shaped DES are not obtained in the homologous case due to the more gradual decrease in the outer density profile with radius and the lower relative velocity of the shock through the faster moving outer portions of the DES. }
\label{fig:radio2}
\end{figure}

Our fiducial model for V1324 Sco corresponds to a fast nova outflow with velocity $v_{\rm w} = 3,000$ km s$^{-1}$, peak mass loss rate $\dot{M}_{w,0} = 4\times 10^{-4}M_{\odot}$ yr$^{-1}$, [fast outflow] duration $\Delta t_{\rm w} = 10$ days (eq.~[\ref{eq:Mdot}]), and total ejecta mass $\approx 4\times 10^{-5}M_{\odot}$.  The DES is assumed to evolve according to the `pulse' model (Fig.~\ref{fig:schematic_density}) with a total mass $M_{\rm DES} = 2\times 10^{-4}M_{\odot}$, characteristic velocity $v_{\rm DES} = 0.32v_{w} \approx 1000$ km s$^{-1}$ and outer radial density index $k = 4$.  A distance $D = 5$ kpc is assumed for V1324 Sco.

Figure \ref{fig:various1} shows various quantities as a function of time since the onset of the optical outburst (June 2, 2012 for V1324 Sco).  The velocity of the central shell $v_{\rm sh}$ remains relatively constant with time at $\sim 2\times  10^{3}$ km s$^{-1}$ as momentum is added by the fast nova outflow but additional mass is swept up from the DES (this is similar to the outflow velocity measured from the optical spectra; \citealt{Ashish+12}).  Initially both the forward and reverse shocks are radiative ($\eta \ll 1$), but with time the reverse shock becomes adiabatic first ($\eta_2 \gtrsim 1$ at $t \sim 30$ days), followed by the forward shock ($\eta_3 \gtrsim 1$ at $t \sim 100$ days).  The temperature behind the reverse shock is high $T_2 \gtrsim 10^{8}$ K at all times, while that behind the forward shock is initially low $\lesssim 2\times 10^{6}$ K before rising to $\sim 5\times 10^{6}$ K at late times.  The lower temperature behind the forward shock $T_{3} \propto (v_{\rm sh}-v_{4})^{2}$ as compared to that of the reverse shock $T_2 \propto (v_{\rm w}-v_{\rm sh})^{2}$ results because the velocity of the forward shock relative to the DES is lower than that of the reverse shock into the fast nova outflow. 

Figure \ref{fig:lum1} shows the power dissipated at the reverse and forward shocks, $\dot{E}_{\rm r}$ (\ref{eq:edotr}) and $\dot{E}_{\rm f}$ (eq.~[\ref{eq:edotf}]), respectively.  The peak power $\dot{E}_{\rm r} + \dot{E}_{\rm f} \gtrsim 10^{38}$ erg s$^{-1}$ is sufficient to account for the luminosity of $\gtrsim 100$ MeV gamma-rays observed by {\it Fermi} (\citealt{Hill+13}; red arrow in Fig.~\ref{fig:lum1}) given an efficiency $\gtrsim 10^{-3}-10^{-2}$ for accelerating hadrons or relativistic electrons at the shock, which is consistent with the range expected from diffusive shock acceleration (e.g.~\citealt{Caprioli&Spitkovsky13}).  

X-rays from the forward and reverse shocks are initially absorbed by the neutral DES, such that a fraction $\epsilon_{\rm opt} = 0.1$ of the shock luminosity is assumed to be re-radiated at optical/UV frequencies ($\S\ref{sec:optical}$).   Figure \ref{fig:lum1} also shows for comparison the V-band light curve\footnote{\url{http://www.aavso.org}} of V1324 Sco with time now normalized such that the optical peak on June 20, 2012 coincides with the time $t = \Delta t_{\rm w}$ of peak mass loss from the white dwarf (eq.~[\ref{eq:Mdot}]; note the difference from our calculations, where this time was defined as $t = 0$).  The qualitative similarity between the shape and magnitude of the predicted luminosity $L_{\rm opt}$ and the V-band light curve suggests that shocks may indeed contribute a substantial fraction of the optical emission of V1324 Sco.  Making a more precise comparison, however, requires an accurate determination of $\epsilon_{\rm opt}$ based on a calculation of the spectrum of the re-processed emission, a task which is beyond the scope of this paper.  It is perhaps worth noting that the UV light curve observed by {\it Swift} UVOT peaked $\sim 14-20$ days after the optical peak (\citealt{Page+12}), as would be expected if a larger fraction of the absorbed shock luminosity escapes in the UV as the UV opacity, which is due primarily to Doppler-broadened metal absorption lines, decreases with time.

At late times the DES becomes optically thin to keV X-rays ($\tau_{\rm \nu_{\rm X},4} \lesssim 1$; eq.~[\ref{eq:taubf}]), soon after which the predicted X-ray luminosity peaks (Fig.~\ref{fig:lum1}).  This transition is predicted to occur earlier for a lower mass fraction $X_Z$ of the intermediate mass CNO elements providing the opacity, resulting in a higher peak luminosity.  For an enrichment $X_Z = 0.01$ as assumed in Figure \ref{fig:lum1}, the X-ray luminosity peaks at $L_X \sim 10^{32}$ ergs s$^{-1}$ on a timescale $\sim 100$ days,  roughly an order of magnitude below the {\it Swift} upper limits.  On the other hand, for gas of higher metallicity $X_Z = 0.1$ similar to that characteristic of nova ejecta (\citealt{Livio&Truran94}), the peak X-ray luminosity is even lower (not shown).  The {\it Swift} 3$\sigma$ upper limits shown in Figure \ref{fig:lum1} are calculated using the Bayesian technique outlined in \citet{Kraft+91}, with the count rates converted into observed luminosities assuming emission from a thermal plasma of temperature $kT =$ 1 keV assuming solar abundances.  The luminosities only correct the count rate for absorption by the ISM based on an assumed column density $N_{\rm H} = 8\times 10^{21}$ cm$^{-2}$ inferred from the optical spectrum (\citealt{Finzell+14}, in prep).

Figure \ref{fig:radio1} (top panel) shows the model predicted radio flux density (eq.~[\ref{eq:Fnu}]) at three frequencies, $4.6$ GHz, $7.4$ GHz, and $36.5$ GHz, for which there are relatively well-sampled light curves for V1324 Sco.  Higher frequencies peak earlier with a somewhat higher flux; the latter behavior can be understood by the flux scaling as $F_{\nu} \propto T_{\nu}^{\rm obs}\nu^{2}R_{\rm sh}^{2}$ (eq.~[\ref{eq:Fnu}]), with the brightness temperature at peak flux (eq.~[\ref{eq:Tobspeak}]) being approximately independent of frequency (eq.~[\ref{eq:Tobspeak}]).  Higher frequencies peak somewhat earlier when the shock radius $R_{\rm sh}$ is smaller, but this is less important than the $\nu^{2}$ dependence in favoring higher frequencies.  In addition to their lower peak flux, the light curves at lower radio frequencies also rise to their peaks more sharply.  These features are all qualitatively consistent with early radio observations of V1324 Sco (\citealt{Finzell+14}, in prep), which are plotted for comparison in Figure \ref{fig:radio1}.  Our fit does a good job of reproducing the 4.6 and 7.4 GHz data near the peak, but the 36.5 GHz data is systematically over-estimated by a factor $\gtrsim 2$.  The model also cannot account for the earliest radio measurements (especially at 4.6 GHz), which appears to require an additional component of ionized gas with a larger radius than that of the forward shock.

The bottom panel of Figure \ref{fig:radio1} shows the spectral indices, $\alpha \equiv d$log $F_{\nu}$/$d$log$\nu$, as defined by the measured flux ratio between the 4.6-7.4 GHz and 7.4-36.5 GHz bands.  At both early and late times, a spectral index $\alpha = 2$ corresponding to optically thick emission is attained.  At early times this optically thick region is the photo-ionized layer ahead of the shock (Region 4), while at late times it represents the temperature of the optically thick central shell (the post shock cooling layer has become radio transparent).  At intermediate times, when the observed emission originates from the hot post shock cooling layer and the flux peaks, the spectral index first rises to a maximum $\gg 2$ before decreasing to a minimum value which (coincidentally) approaches the flat value $\alpha \approx 0$ predicted for single temperature, optically thin free-free emission.  At late times the cooled shell expands with a uniform minimum temperature $\sim 10^{4}$ K and relatively constant velocity, after which point the evolution can be matched smoothly onto the standard scenario of ballistically expanding photo-ionized gas (\citealt{Seaquist&Bode08}). 

A `pulse' profile was adopted as our fiducial model for the DES in fitting to V1324 Sco instead of the homologous model because we found that the latter was unable to account for the fast rise and narrow width of the radio light curve.  To illustrate this point, in Figure \ref{fig:radio2} we show the predicted radio emission for the homologous model with otherwise identical parameters for the properties of the fast nova outflow and the DES, i.e. we assume the same characteristic velocity $v_{\rm DES} = 3,000$ km s$^{-1}$ and same outer density index $k = 4$ as in the pulse model shown in Figure \ref{fig:radio1}.  Figure \ref{fig:radio2} shows that the sharply peaked 4.6 and 7.4 GHz light curves (of duration $\delta t \ll t$) are clearly not reproduced by the homologous model, even for a steep radial index $k =4$ (larger values of $k$ did not improve the fit).  This slower evolution is due to the more gradual decline in density with radius in the homologous model, and the decreasing relative velocity between the shell and the DES $v_4 \propto r$ (eq.~[\ref{eq:homovel}]).  

\subsubsection{Range of Models}
\label{sec:range}

We have explored a range of pulse models by varying the properties of the nova outflow ($\dot{M}_0, v_{\rm w}, \Delta t_{\rm w}$) and the DES ($M_{\rm DES},  v_{\rm DES}, k$) while fixing the DES duration $\Delta t = 18$ days to that assumed for V1324 Sco.  Our results are summarized in Table \ref{table:results}.  

\begin{figure}
\includegraphics[width=0.5\textwidth]{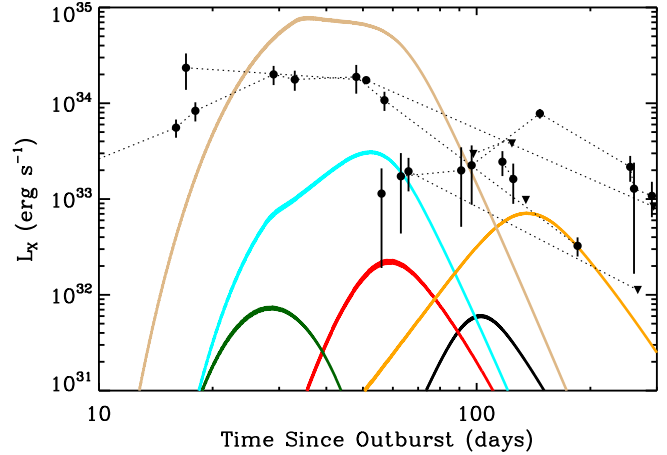}
\caption{X-ray light curves for several models listed in Table \ref{table:results}.  X-ray detections ({\it circles}) and upper limits ({\it triangles}) from \citet{Mukai+08} are shown for comparison in black, with data from the same novae connected by dashed lines.  The Nova shown include V838 Her; V1974 Cyg; V351 Pup; V382 Vel; N LMC 2000; V4633 Sgr; NLMC 2005; V5116 Sgr; V1663 Aql; V1188 Sco; V477 Sct; V476 Sct; V382 Nor.}
\label{fig:Xlums}
\end{figure}

Figure \ref{fig:Xlums} shows the range of $\sim$ 1 keV X-ray light curves for the models in Table \ref{table:results} for assumed CNO mass fraction $X_Z = 0.01$.  For comparison we show measurements or upper limits from {\it ROSAT} and {\it Swift} for 13 novae as compiled by \citet{Mukai+08}.  Most of the predicted luminosities are consistent with the peak luminosities and durations of the observed X-ray light curves, while others are below the detections, consistent with the upper limits.  The X-ray luminosities predicted by our model thus appear generally consistent with observations given reasonable variations in the properties of the fast nova outflow and the DES.

\begin{figure}
\includegraphics[width=0.5\textwidth]{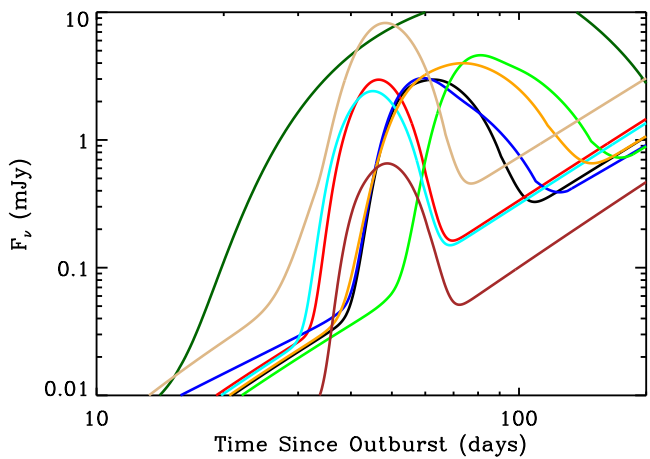}
\caption{Radio flux densities at 7.6 GHz for several models listed in Table \ref{table:results} and shown in Figure \ref{fig:Xlums}.
}
\label{fig:radiolums}
\end{figure}

Figure \ref{fig:radiolums} shows examples of the 7.4 GHz light curves for the same models shown in Figure \ref{fig:Xlums}.  A variety of peak times and fluxes are obtained, but in all cases the same qualitative feature of a `bump' above the otherwise $\sim 10^{4}$ K photosphere is clearly present.  The qualitative features presented for V1324 Sco may thus be generic to shock interaction in novae.


\begin{table*}
\begin{scriptsize}
\begin{center}
\vspace{0.05 in}\caption{Summary of Nova Shock Models$^{\dagger}$}
\label{table:results}
\begin{tabular}{ccccccccccc}
\hline \hline
\multicolumn{1}{c}{color$^{(a)}$} &
\multicolumn{1}{c}{$\dot{M}_{\rm w,0}$$^{(b)}$} &
\multicolumn{1}{c}{$\Delta t_{\rm w}$$^{(c)}$} & 
\multicolumn{1}{c}{$v_{\rm w}$$^{(d)}$} & 
\multicolumn{1}{c}{$M_{\rm DES}$$^{(e)}$} &
 \multicolumn{1}{c}{k$^{(f)}$} &
 \multicolumn{1}{c}{$v_{\rm DES}/v_w$$^{(g)}$} &
\multicolumn{1}{c}{$t_{\rm X}^{\rm peak}$$^{(h)}$} & 
\multicolumn{1}{c}{$L_{\rm X}^{\rm peak}$$^{(i)}$} &
\multicolumn{1}{c}{$t_{7.6{\rm GHz}}^{\rm peak}$$^{(j)}$} & 
\multicolumn{1}{c}{$F_{7.6 {\rm GHz}}^{\rm peak}$$^{(k)}$}  \\
\hline
 - & ($M_{\odot}$ yr$^{-1}$) & (d)  &  ($10^{3}$ km s$^{-1}$) & ($M_{\odot}$) & - & (cm) & (d) & ($10^{33}$ erg s$^{-1}$) & (d)  & (mJy)\\
\hline 
\\
black & $4\times 10^{-4}$ & 10 & 3 & $2\times 10^{-4}$  & 4 & 0.32 & 100 & 0.06 & 63 & 3.0 \\
red & $1\times 10^{-3}$ & - & - & -  & - & - & 60 & 0.2 & 47 & 3.0 \\
blue & $4\times 10^{-4}$ & 5 & - & -  & - & - & `60 & 0.0 & 59 & 3.0 \\
green & - & 10 & - & $4\times 10^{-4}$ & - & - & 210 & 4$\times 10^{-4}$ & 81 & 4.6 \\
cyan & - & - & - & $1\times 10^{-4}$ & - & - & 52 & 3.1 & 45 & 2.4 \\
orange & - & - & - & $2\times 10^{-4}$ & 3 & - & 136 & 0.7 & 74 & 4.0 \\
dark green & - & - & - & - & 4 & 0.6 & 29 & 0.07 & 100 & 15 \\
fuscia & - & - & - & -  & - & 0.15 &  158 & 2$\times 10^{-5}$ & 49 & 0.65 \\
gray & - & - & 5 & -  & - & 0.32 &  36 & 77 & 48 & 8.2 \\

 \\
  \\
\hline
\hline
\end{tabular}
\end{center}
$^{\dagger}$ All models are calculated assuming a `pulse' profile for the density/velocity structure of the DES (Fig.~\ref{fig:schematic_density}) and assuming a delay $\Delta t = 18$ days between the DES and the peak in the mass loss rate of the fast wind, as characterized by the delay between the initial optical rise and optical peak of V1324 Sco.  $^{(a)}$Color coding of models in Figs.~\ref{fig:Xlums} and \ref{fig:radiolums}. $^{(b)}$Peak mass loss rate of the white dwarf (see eq.~[\ref{eq:Mdot}]). $^{(c)}$Characteristic duration of peak mass loss from white dwarf. $^{(d)}$Velocity of the fast nova outflow. $^{(e)}$Total mass of external shell. $^{(f)}$Index of density profile of external shell.  $^{(g)}$Characteristic velocity of the DES relative to the fast outflow.  $^{(h)}$Time of peak X-ray luminosity measured from onset of optical outburst. $^{(i)}$Peak X-ray luminosity.  $^{(j)}$Time since outburst of first radio peak at 7.4 GHz measured from onset of optical outburst.  $^{(k)}$Peak 7.4 GHz flux (for an assumed distance of 5 kpc).
\end{scriptsize}
\end{table*}

\section{Discussion}
\label{sec:discussion}

Our results support the conclusion that shock interaction plays an important role in shaping observational features of novae across the electromagnetic spectrum. 

The early optical emission of novae is well modeled as originating from an optically-thick photosphere, corresponding to the stationary atmosphere of the white dwarf or an expanding wind (e.g.~\citealt{Hauschildt+97}).  This emission is traditionally thought to be powered by thermal energy generated by nuclear burning on the white dwarf surface (e.g.~\citealt{Prialnik86}).  Here we have shown, however, that the absorption and re-emission of X-rays from nova shocks by neutral gas ahead of the shock can power optical luminosities of a similar magnitude and duration to observed nova emission (Fig.~\ref{fig:lum1}).  The possibility that shock dissipation contributes to the optical light curve is supported by the temporal coincidence between the optical and gamma-ray peak in the majority of the {\it Fermi-}detected novae (\citealt{Hill+13b}), as the gamma-ray emission is undoubtedly shock powered.  Determining the contribution of shocks versus nuclear burning to nova light curves will require more detailed predictions for the spectrum of the shock-powered emission, which is however beyond the scope of this work.

Although most novae decline monotonically in visual brightness after their initial peak, a small fraction achieve a secondary peak (or peaks) in brightness (e.g.~\citealt{Strope+10}; \citealt{Kato&Hachisu11}).  Our results suggest the intriguing possibility that such secondary maxima could be shock powered, even if the first peak is powered by emission from the white dwarf surface.  Variability of the optical light curve in such a scenario can then be understood as due to either variations in the nova outflow or variations in the DES density: a complex light curve may therefore suggest, but would not necessarily require, a complex mass loss history from the white dwarf.


Flat radio spectra at early times have previously been used to argue for a synchrotron origin for nova radio emission (e.g.~\citealt{Anupama&Kantharia05}; \citealt{Sokoloski+08}).  Our results show that a nearly flat spectrum can also be explained as thermal radiation from the shock (bottom panel, Fig.~\ref{fig:radio1}) without the need to invoke non-thermal radiation or relativistic particle acceleration (\citealt{Krauss+11}; \citealt{Weston+13}).  A testable prediction of our model is that the spectral index is initially {\it higher} than the optically thick value during the early radio rise and that it varies strongly with time.  A related prediction is that the radio light curve is characterized by a faster rise to maximum, and a later time to peak, at progressively lower frequencies.

The gamma-rays from some novae provide clear evidence for non-thermal processes in nova shocks (e.g.~\citealt{Abdo+10}; \citealt{Martin&Dubus13}).  Our model fit to V1324 Sco requires that $\gtrsim 10^{-3}$ of the shock luminosity must be placed into non-thermal electrons or hadrons to explain the observed gamma-ray luminosity (Fig.~\ref{fig:lum1}), an acceleration efficiency that appears to be consistent with theoretical models for diffusive particle acceleration at non-relativistic shocks (e.g.~\citealt{Caprioli&Spitkovsky13}).  Future work (Paper II) will explore the implications of non-thermal radiation for the multi-wavelength picture of nova shocks, including aspects neglected here such as the effects of magnetic fields and cosmic rays on the compression of the post-shock cooling layer.  

The characteristic luminosities and timescales of $\sim 1$ keV X-rays predicted by our model are consistent with those observed in the current sample of novae (Fig.~\ref{fig:Xlums}).  The luminosities predicted by our fiducial model are also consistent with the {\it Swift} X-ray upper limits for V1324Sco (Fig.~\ref{fig:lum1}) for essentially all physical values of the CNO mass fraction of the DES $X_Z \gtrsim 0.01$.  In some novae X-rays are observed simultaneous with the radio peak (\citealt{Krauss+11}), while in others there does not appear to be any X-rays at times coincident with the radio rise; our model can accommodate both possibilities (compare Figs.~\ref{fig:Xlums} and \ref{fig:radiolums}).  Finally, if the optical light curve is indeed powered by the reprocessing of X-rays from the shock, then a steep drop in the optical light curve may be accompanied by a simultaneous X-ray brightening. 

The density of the DES at the time it collides with the fast wind (eq.~[\ref{eq:nin}]) is similar to those characterizing the ambient medium in symbiotic systems produced by the wind of the giant companion.  The observed X-ray luminosities due to shocks in symbiotic novae (e.g.~\citealt{Sokoloski+06}; \citealt{Nelson+12d}) are however considerably higher than those predicted here for classical nova outflows.  This difference results from (1) the potentially higher metallicity (and hence X-ray opacity) of the shocked gas in the classical nova case if it originates from the white dwarf instead of the giant companion; and  (2) the lower temperature of the forward shock as results from the lower {\it relative} velocity between the fast nova outflow and the slower DES, as compared to the case when the fast nova outflow collides with the effectively stationary stellar wind of the red giant.

Our model for V1324 Sco suggests that the DES and the fast nova wind possess a combined mass $\sim$ few $\times 10^{-4}M_{\odot}$.  This value is consistent with that inferred based on fitting the second late radio peak to the standard model for expanding thermally ionized plasma (\citealt{Finzell+14}, in prep).

The physical source of the DES in classical novae remains an important open question.  \citet{Williams12} proposes that novae with `Fe II' spectra are formed in a large circumnova envelope of gas produced by the secondary star.  Potential sources of DES in such a scenario include a common envelope phase prior the outburst (\citealt{Shankar+91}) or a circumnova reservoir of gas that is produced by the mass loss through the outer L3 Lagrangian point (e.g.~\citealt{Sytov+07}).  If the DES does not originate from the white dwarf, then it must require some additional source of acceleration, such as radiation pressure from the nova outburst (e.g.~\citealt{Williams72}) or by orbital energy injected by the binary (\citealt{Shankar+91}), to achieve the observed velocities.  


If we instead assume that the DES originates from the white dwarf surface starting near or soon after the initial optical outburst, then it is not necessarily surprising that the `pulse' model provides a better fit to the radio data of V1324 Sco than homologous expansion.  This is because by the time of the early radio peak at $t \sim 70$ days, the DES has traveled a distance $\sim  v_{\rm DES}t$ that is only a factor of $\sim 4$ times larger than its assumed width $v_{\rm DES}\Delta t$, in which case its initial shape is maintained if the DES possesses a velocity dispersion $\Delta v_4/v_{\rm DES} \lesssim 0.25$ which is consistent with the small measured velocity dispersions $\sim 35-400$ km s$^{-1}$ of the narrow-line absorbing gas in classical novae (e.g.~\citealt{Williams08}).  It remains to be seen if such low velocity dispersions are obtained by realistic physical models for nova outflows (e.g.~\citealt{Yaron+05}).  

If the profile of the shocked DES indeed retains some memory of its initial conditions, then the steep outer density profile ($k = 5$) needed to fit the rapid radio peak provides an important clue to its origin.  Such a steep profile requires an abrupt onset to the initial stage of mass loss from the white dwarf.  Hydrodynamical models of classical nova outbursts in fact show that the thermonuclear runaway (TNR) produces a shock wave, which moves through the envelope of the white dwarf and accelerates the surface layers to a high velocity (e.g.~\citealt{Prialnik86}).  In the interpretation that such a shock produces a steep outer density profile, then the sharp early radio maximum serves as an indirect imprint of the TNR event.  Future work should compare the density profile predicted by theoretical models for the early nova outflows to those we find are required to explain the rapid radio evolution of V1324 Sco.

Our main conclusion, that many aspects of nova emission can in principle be explained by thermal emission from shocks, comes with several important caveats.  First, our model assumes a one-dimensional geometry, despite a multitude of observations which suggest that nova outflows are intrinsically multi-dimensional (e.g.~\citealt{OBrien&Cohen98}; \citealt{Porter+98}; \citealt{Sokoloski+13}; \citealt{Ribeiro+13}).  A 1D model should still provide a relatively accurate description of the shock emission if the DES subtends a substantial fraction of the solid angle surrounding the white dwarf, as would be the case if it possesses, for instance, a toroidal geometry.  Such a geometry would allow the faster nova outflow to emerge unhindered along the pole, possibly helping to explain the low frequency radio excess observed in V1324 Sco at the earliest epochs (Fig.~\ref{fig:radio1}; such emission is not easily accommodated by our model due to the small radius of the blast-wave at such early times).  

Secondly, our model for the radio assumes that free-free emission instead of line cooling dominates the post-shock cooling down to temperatures $\sim 10^{6}$ K characterizing the radio photosphere (eq.~[\ref{eq:Tobspeak}]).  This assumption is justified by the argument that UV/X-rays from the forward shock may penetrate and over-ionize the post-shock gas, suppressing line cooling ($\S\ref{sec:postshock}$).  Such over-ionization would be aided by the presence of non-thermal pressure support behind the shock, such as due to shock-accelerated cosmic rays or magnetic fields, which would act to reduce the density (and hence the rate of radiative recombination) at the radio photosphere.  If such over-ionization is insufficient to suppress line cooling, this would substantially reduce the brightness temperature of the post-shock layer (eq.~[\ref{eq:T0nu}]) to a value incompatible with the measured flux of the early radio bump.  The only remaining explanation in this case is that synchrotron radiation from electrons accelerated by the forward shock, instead of the thermal emission considered here, is responsible for powering the early radio emission.

\section{Conclusions}
\label{sec:conclusions}

We have developed a model for the shock interaction between the fast outflows from nova eruptions and a `dense external shell' or DES, which may represent an earlier episode of mass loss from the white dwarf.  This interaction is mediated by a forward shock driven ahead into the DES and a reverse shock driven back into nova ejecta.  Our results are summarized as follows:
\begin{itemize}
\item{Shocks heat the gas to X-ray temperatures (Fig.~\ref{fig:schematic}).  The shocks are radiative at early times, producing a dense cooling layer downstream, before transitioning to become adiabatic at late times.  Cooled gas swept up by the shocks accumulates in a central shell, the inertia of which largely controls the blast-wave dynamics.}
\item{X-ray/UV photons from the forward shock can penetrate the downstream cooling layer ($\S\ref{sec:postshock}$).  By photo-ionizing the gas and reducing the neutral fraction below that in collisional ionization equilibrium, we speculate that line cooling will be suppressed, in which case free-free emission dominates cooling of the post-shock gas to lower temperatures than is usually assumed.}
\item{At early times when the forward shock is radiative, radio emission originates from a dense cooling layer immediately downstream of the shock (Fig.~\ref{fig:layer_schematic}).  The predicted radio maximum is characterized by sharper rises, and later peak times, at progressively lower frequencies.  The brightness temperature at peak flux is approximately independent of frequency at a characteristic value $\sim 10^{6}$ K (eq.~[\ref{eq:T0nu}]).}
\item{X-rays from the forward shock are absorbed by the neutral DES at early times.  Re-radiation of this luminosity at optical/UV frequencies may contribute appreciably to the optical light curves of nova.}
\item{At late times when the absorbing column of the DES ahead of the shock decreases, X-rays escape to the observer.  The predicted X-ray luminosities peak at values $\lesssim 10^{32}-10^{34}$ erg s$^{-1}$ on timescales of weeks to months, consistent with observations of classical novae.}  
\item{Our model provides an adequate fit to the radio data of V1324 Sco for reasonable assumptions about the properties of the nova outflow motivated by the optical observations, if one assumes a DES with a velocity $\sim 1,300$ km s$^{-1}$ and mass $\sim 2.5\times 10^{-4}M_{\odot}$.  The total ejecta mass (fast nova outflow + DES) of $\sim$ few $\times 10^{-4}M_{\odot}$ is consistent with that inferred independently by modeling the late radio emission as uniformly expanding photo-ionized gas (\citealt{Finzell+14}, in prep).}
\item{The sharp early peak in the radio light curve of V1324 Sco requires a steep outer radial density profile for the DES, which may provide evidence for the rapid onset of mass loss from the white dwarf following the thermonuclear runaway.}
\end{itemize}

\section*{Acknowledgements}
BDM acknowledges conversations with Bob Williams.  We thank Jennifer Weston, Thomas Finzell, and Yong Zheng for helping to reduce and analyze the radio observations of V1324 Sco.  The National Radio Astronomy Observatory is a facility of the National Science Foundation operated under cooperative agreement by Associated Universities, Inc.  BDM acknowledges support from the Department of Physics at Columbia University.  JLS acknowledges support from NSF award AST-1211778.



\end{document}